\RequirePackage{fix-cm}
\documentclass[smallextended]{svjour3}       % onecolumn (second format)
\smartqed  % flush right qed marks
\usepackage{natbib}
\usepackage{graphicx}
\usepackage{xspace}
\usepackage{nicefrac}
\usepackage{caption}
\usepackage{subcaption}
\usepackage{amsfonts}
\usepackage{amsmath}

\newcommand{\binvec}{\ensuremath{\mathcal{S}}\xspace}
\renewcommand{\vec}[1]{\ensuremath{\mathbf{#1}}}
\DeclareMathOperator*{\argmax}{arg\,max}

\newcommand{\out}{\ensuremath{\vec{x}}\xspace}

\usepackage{tikz}
\usetikzlibrary{matrix}
\usetikzlibrary{arrows,fit}
\usetikzlibrary{positioning}
\usetikzlibrary{calc}
\usetikzlibrary{shapes,decorations}
	\tikzset{
		inner sep=0pt, outer sep=0pt, minimum size=0pt, thick,
		level/.style={sibling distance = (\columnwidth/16)*2^(4-#1)},
		winner/.style={minimum size=1.5em, circle, draw, fill=white, font={\footnotesize}},
		split/.style={minimum size=1.5em, inner sep=1pt, circle split, draw, fill=white, font={\tiny}},
		leaf/.style={inner sep=.15em, font={\footnotesize}},
		ball/.style={minimum size=.4em,circle,fill=black},
		beats/.style={thick,->,>=stealth',draw},
		tline/.style={thick,draw}
	}

\begin{document}

\title{Flexible Representative Democracy
}
\subtitle{An Introduction with Binary Issues}

\titlerunning{Flexible Representative Democracy}        % if too long for running head

\author{ Ben Abramowitz \and Nicholas Mattei }


\institute{Ben Abramowitz \at
              Tulane University \\
            %   Tel.: +123-45-678910\\
            %   Fax: +123-45-678910\\
              \email{babramow@tulane.edu}           %  \\
           \and
           Nicholas Mattei \at
              Tulane University \\
              \email{nsmattei@tulane.edu}
}

\date{Received: date / Accepted: date}

\maketitle

\begin{abstract}
% What
We introduce Flexible Representative Democracy (FRD), a novel hybrid of Representative Democracy (RD) and Direct Democracy (DD) in which voters can alter the issue-dependent weights of a set of elected representatives.
% Why
In line with the literature on Interactive Democracy, our model allows the voters to actively determine the degree to which the system is direct versus representative. However, unlike Liquid Democracy, Flexible Representative Democracy uses strictly non-transitive delegations, making delegation cycles impossible, and maintains a fixed set of accountable, elected representatives.
% How
We present FRD and analyze it using a computational approach with issues that are binary and symmetric. We compare the outcomes of various voting systems using Direct Democracy with majority voting as an ideal baseline.
% Results
First, we demonstrate the shortcomings of Representative Democracy in our model. We provide NP-Hardness results for electing an ideal set of representatives, discuss pathologies, and demonstrate empirically that common multi-winner election rules for selecting representatives do not perform well in expectation.
%Theory
To analyze the effects of adding flexibility, we begin by providing theoretical results on how issue-specific delegations determine outcomes.
%Empirical
Finally, we provide empirical results comparing the outcomes of Representative Democracy, proxy voting with fixed sets of proxies across issues, and Flexible Representative Democracy with issue-specific delegations. Our results show that variants of Proxy Voting yield no discernible benefit over unweighted representatives and reveal the potential for Flexible Representative Democracy to improve outcomes as voter participation increases.
% Include keywords, PACS and mathematical subject classification numbers as needed.
\end{abstract}

\begin{acknowledgements}
Portions of this work were completed while both Nicholas Mattei and Ben Abramowitz were at IBM Research, Yorktown Heights, New York. Ben Abramowitz was supported by the CIFellows Project. Nicholas Mattei was supported in part by NSF Awards IIS-RI-2007955, IIS-III-2107505, and IIS-RI-2134857, as well as an IBM Faculty Award and a Google Research Scholar Award.

This material is based upon work supported by the National Science Foundation under Grant \#2030859 to the Computing Research Association for the CIFellows Project.
\end{acknowledgements}

\section{Introduction}\label{sec:intro}
% Since the Athenian \emph{Ecclesia} in 595 BCE Direct Democracy (DD) as an ideal collective decision making scheme has loomed large in the western imagination~\cite{dunn1995democracy}. 
Direct Democracy becomes impractical at scale because it places too much burden on individual decisions makers: everyone must be well-informed on every issue and always available to vote~\citep{green2015direct}. In addition to the attention requirements, voters are also required to know and be able to articulate their preferences at the time of every vote. While preferences and preference learning are large research areas in AI~\citep{DHKP11a,FuHu10a}, every voter may not have enough knowledge, information, time, or energy to participate, particularly when issues are complex.

Given the prohibitive costs of implementing a large-scale Direct Democracy, we often resort to forms of representation, relying upon a set of proxies to make decisions on the voters' behalf. 
Sets of representatives are used in many contexts to reduce the computation and communication burden of decision makers. Countries have parliaments, companies have shareholders, and even groups of robotic agents select leaders to represent them~\citep{Yu:2010:CDM:1838186.1838192}. 

It is often beneficial to elect fixed committees which meet certain axiomatic criteria. For example, committees should be proportional and have justified representation of the voters~\citep{aziz2017justified}, or satisfy other forms of representation~\citep{anshelevich2021representative}. Intuitively, the difficulties in electing committees carry through to the setting of Representative Democracy (RD) where the committee makes decisions in the interest of the voters/agents who elect them~\citep{skowron2015we}. 

Within computer science, many applications face the task of selecting representatives for downstream decision making. In portfolio selection, a particular set of algorithms and hyper-parameters are selected from a large pool of candidates and then used as representatives for later problems~\citep{khudabukhsh2009satenstein}. In multi-agent systems, the role assignment problem uses distributed voting to decide on tasks for agents~\citep{zhu2012group}. And in group recommendation settings, elections correspond to picking a set of experts to later make decisions. The Computational Social Choice community has produced a large body of research on how to select and weight representatives~\citep{BCELP16a}. Indeed, using multi-winner voting~\citep{SFL16a}, we can view the winners as a set of exemplars that may be used to decide some downstream application -- e.g., we select a set of points in space and then aggregate these points (votes) over the set.

Since Direct Democracy is impractical and Representative Democracy comes with inherent trade-offs and limitations, hybridizations of the two have arisen under the umbrella of \emph{interactive democracy}. This idea, coupled with modern communication technologies, has spawned a large number of proposed democratic decision making systems, and interactive democracy has become an important area of research and application for AI~\citep{brill2018interactive}. Liquid Democracy in particular has received significant attention in the political science~\citep{green2015direct}, AI~\citep{kahng2021liquid, golz2021fluid} and agents communities~\citep{brill2018pairwise}, and has been implemented in both corporate~\citep{hardt2015google} and political settings~\citep{blum2016liquid,behrens2014principles}.
As the security of internet applications and accessibility improve, many scholars and companies are turning toward computer systems to address issues with democratic decision making systems; some going so far as to suggest that we create an AI-based Direct Democracy or an ``Augmented Democracy" \footnote{https://www.peopledemocracy.com/}.

In contrast to existing interactive democracy proposals, the voting systems within the class of mechanisms we define as Flexible Representative Democracy maintain a set of expert representatives while allowing voters to guarantee their own representation without raising the minimum required burden on them. In an FRD voters elect a set of representatives to serve a term during which make decisions. Each voter, by default, allocates some (potentially zero) fraction of their voting power to each representative. If this default allocation is uniform and we stop here, we are left with a traditional representative body where each representative has equal power. However, for each issue under consideration in FRD, the voters can deviate from this default by delegating their voting power over any subset of representatives. If all voters use their option to delegate on each issue, as long as there is at least one representative who agrees with each voter's view, the outcome can become exactly that of Direct Democracy. Voters have both the election \emph{and} the flexible delegation option as tools for achieving representation and holding representatives accountable.

In a Flexible Representative Democracy, voters have great flexibility in determining how they are represented. 
% When applicable, the mandated disclosure of representatives' votes guarantees that an attentive voter \emph{can} be fully informed about how their voting power will be and was used. 
For example, the day after the election an inattentive voter might choose a few elected representatives they trust, apportion the power of their vote to these few for all future issues, and pay no attention until the next election. A more attentive voter might alter their allocations on an issue-by-issue basis as issues arise, reacting to representatives' deliberations and pronouncements. In general, voters determine the granularity with which they privately express their preferences over issues via the representatives. Thus, the degree to which Direct Democracy is emulated by a Flexible Representative Democracy depends both on the caliber of the representatives and the fastidiousness of the voters.

\subsection{Contributions} We introduce Flexible Representative Democracy (FRD), a new class of voting systems with the ability to transition smoothly, at the discretion of the voters, between Representative Democracy and Direct Democracy. Our proposal for FRD solves standing issues in the literature on interactive democracy including maintaining a fixed, elected committee to generate legislation and making delegation cycles impossible. We analyze our model in decision making scenarios involving binary, symmetric issues and (1) show that electing an optimal set of representatives is hard for any large-scale Representative Democracy that uses a multi-winner voting rule, (2) investigate the performance of various deterministic multi-winner voting rules to select representatives, (3) demonstrate the theoretical ability of issue-specific delegations under FRD to overcome the limitations of Representative Democracy, and (4) provide empirical results demonstrating that FRD outperforms both Representative Democracy and Proxy Voting in terms of emulating Direct Democracy.

\section{Preliminaries}\label{sec:prelims}

\paragraph{Direct Democracy}\label{sec:prelims_DD}
In a Direct Democracy, every voter can express their preferences directly on every issue to be decided upon. Their preferences are then aggregated to select an outcome for each issue.
Let $V$ be a set of voters and $S$ a set of issues. Voters express their preferences for each issue $s \in S$. The collection of voter preferences regarding $s$ is denoted $P_V^s$, and the collective profile of preferences over the issues as a whole is $P_V$.
We denote by $R$ the \emph{decision rule} that aggregates voter preferences into outcomes. For brevity, we assume that the same decision rule $R$ is applied separately for all issues. The outcome on issue $s$ is $R(P_V^s)$, and the vector of outcomes is denoted by the shorthand $R(P_V) = (R(P_V^1), \ldots, R(P_V^{|S|}))$.

\paragraph{Representative Democracy}\label{sec:prelims_RD}
In a Representative Democracy, the voters do not report their preferences regarding each issue. Instead, the voters elect a set of representatives who then vote on behalf of the whole population for each issue.
Initially, there is a slate of candidates $C$ who vie for election. The voters express preferences over the candidates, collectively denoted by the profile $P_{VC}$. An \emph{election rule} $E$ then selects a subset of candidates as representatives: $E(P_{VC}) = D \subseteq C$.
Once elected, the representatives cast their votes on each issue in a set $S$, forming profile $P_D = (P_D^1, \ldots, P_D^{|S|},)$.
Finally, the set of outcomes is determined by applying a decision rule $R(P_D) = (R(P_D^1), \ldots, R(P_D^{|S|}))$.

\paragraph{Weighting Representatives}\label{sec:prelims_weights}
Many decision rules treat the votes from different agents similarly. In other words, the decision rule are \emph{anonymous} in that it does not matter which agent each vote comes from.
One way to relax anonymity is to weight the agents and use a weighted decision rule.
Our focus is on voting systems that weight representatives.
A rep could have the same weight across all issues or a different weight for each issue.
Letting $W$ represent the weighting of the representatives, the decision outcomes are computed individually $R(P_D^s, W^s)$, with the vector of outcomes denoted in shorthand by $R(P_D, W)$.
%
% When $R$ is a weighted decision rule, $R(P_V)$ and $R(P_D)$ correspond to the cases where all voters or reps are given the same weight, respectively.
% %
% For example, when $R(P,W)$ is the weighted majority rule, $R(P)$ is simple majority rule.

\paragraph{Electoral Weighting}\label{sec:prelims_WRD}
One way to weight representatives is based on the electoral profile, i.e. $R(P_D, P_{VC})$.
We assume there is a subroutine of $R$ that incorporates $P_{VC}$ by weighting the reps.
Specifically, we take there to be some function $f$ that maps electoral profiles to weights: $f(P_{VC}) = W$, and these weights are used to compute the decision outcomes $R(P_D, W)$.
In this sort of weighting scheme, the weights of reps remain constant across issues.

\paragraph{Proxy Voting}\label{sec:prelims_proxy}
In proxy voting, there is a set of ``reps" $D$ available as potential proxies for each voter to choose from. The proxies may or may not be elected. Each voter selects a single rep as their proxy, and the proxies are given weights based on how many voters chose them.
As with RD, we will identify a subroutine $f$ as translating proxy choices into weights.
If we denote the choices of the voters regarding proxies as profile $P_{VD}$, then with some weighting function $f(P_{VD}) = W$, the outcomes are $R(P_D, W)$.

Proxy voting can be generalized by allowing each voter to select more than one proxy or assigning scores to the proxies.
Unlike RD with electoral weighting, in proxy voting the weights are independent of how the set of reps $D$ is constructed, and depend instead on preferences over the reps expressed after the set of reps is determined. Also, proxy voting is frequently used on an issue-by-issue basis, where voters can report $P_{VD}^s$ for each issue $s$ rather than using a single weighting of representatives across all issues.

\paragraph{Abstention}\label{sec:prelims_abstention}
In reality, agents do not always participate in every aspect of the decision-making process available to them. Individual voters may not cast their votes in DD, may not express their preferences over candidates in RD, and may not select a proxy under proxy voting. Similarly, reps and proxies may not cast votes.
Each decision rule must have a way of dealing with such abstentions.
In our model, abstentions are accounted for by $f$ when computing $W$. This keeps the definition of $R$ simple.

\paragraph{Flexible Representative Democracy}\label{sec:prelims_FRD}
In a Flexible Representative Democracy, the voters begin by electing a set of representatives $D = E(P_{VC})$. Voters then have the option to express preferences  over the elected representatives, e.g. choose proxies, on an issue-by-issue basis ($P_{VD}$). The outcome of each issue is then determined by $R(P_D^s, W^s)$ where $W^s = f(P_{VC}, P_{VD}^s)$ for some weighting protocol $f$. Once again, the vector of outcomes across the issues is denoted $R(P_D, W)$.

While we do not address the temporal aspects of the voting systems in our work, one feature of Flexible Representative Democracy to keep in mind is that if the reps fix their votes publicly on each issue before the decision rule is applied, this affords the voters an opportunity to update $P_{VD}^s$ with knowledge of $P_D^s$ before each decision is made.

% \begin{definition}[Agreement]
% Given a decision rule $R$, the representatives \emph{agree} with the voters on issue $s$ if $R(P^s_V) = R(P^s_D)$.
% \end{definition}

% \begin{definition}[Coverage]
% Given a decision rule $R$ and voter profile $P_V$, we say that the representatives \emph{cover} an issue $s$ if $\exists D' \subseteq D$ such that $R(P^s_{D'}) = R(P^s_V)$
% \end{definition}

% \begin{definition}[Full Coverage]
% Given a decision rule $R$ and voter profile $P_V$, we say that the representatives \emph{fully cover} an issue $s$ if $\forall V' \subseteq V$, $\exists D' \subseteq D$ such that  $R(P^s_{D'}) = R(P^s_{V'})$.
% \end{definition}

\subsection{Binary Issues Model}\label{sec:prelims_model}
Let $V = \{v_1, \ldots, v_n\}$ be an indexed set of voters, and $S = \{s^1, \ldots, s^r\}$ be an indexed set of binary issues. 
The two possible outcomes for each issue are denoted $\{0,1\}$. 
On each issue $s^i$, each voter $v_j$ has a preference $v_j^i \in \{0,1\}$.
The full preferences of voter $v_j$ across the issues is denoted $\vec{v}_j = \{v_j^1, \ldots, v_j^r\}$ and $P_V = \{\vec{v}_j : v_j \in V\}$ denotes the profile of voters' preferences as a whole.
When $R$ is majority rule with random tie-breaking, i.e. coin flip, we label the alternatives such that $R(P_V^s)=1$ without loss of generality.

Let $D = \{d_1, \ldots, d_k\}$ be an indexed set of representatives with preferences $\vec{d}_l = \{d_l^1, \ldots, d_l^r\} \in \{0,1\}^r$ comprising the preference profile $P_D = \{\vec{d}_l : d_l \in D\}$. We will assume that the number of representatives $k$ is always odd to avoid unnecessary tie-breaking.

% \begin{definition}[Full Coverage]
% A set of representatives $D$ fully covers an issue $s^i$ if $\exists d_l \in D$ such that $d_l^i = 1$ and $\exists d_l \in D$ such that $d_l^i = 0$.
% \end{definition}

We hold Direct Democracy with majority voting and full participation as our ideal for comparison. Therefore the ideal outcome is $\vec{1}$.
We will be evaluating the performance of FRD and other voting systems in terms of the fraction of issues on which they produce the outcome corresponding to the voter majority.

\begin{definition}[Agreement]
Let \binvec be the set of all binary vectors of length $r$. For any two vectors $\vec{x}$, $\vec{y} \in \binvec$, we measure their similarity or \emph{agreement} $L(\vec{x}, \vec{y})$ by the fraction of their entries that are the same.\footnote{Recall that the order of the binary issues does not matter in our model and the issues are treated identically, so other common measures of distance/similarity used in the Social Choice literature are either equivalent to the Hamming Distance or not applicable.} In other words, the agreement between two vectors is inverse to the normalized Hamming Distance between them: $$L(\vec{x}, \vec{y}) = 1 - \frac{1}{r} \sum\limits_{i=1}^r |x^i - y^i|$$
\end{definition}

If $\out$ is a vector of outcomes yielded by some voting procedure, we measure its \emph{agreement} with the voter majority $L(\out, \vec{1})$.

\begin{definition}[Majority Agreement]
The majority agreement between a set of representatives $D$ and the set of voters $V$ is the fraction of issues on which the majority of reps agrees with the voter majority, i.e. $L(R(P_V), R(P_D)) = \frac{1}{|S|} |\{s^i \in S : \sum\limits_{d_l \in D} d_l^i > \frac{|D|}{2}\}|$ where $R$ is majority rule.
\end{definition}

We also define a weaker notion than majority agreement that we call coverage. Majority agreement on any issue implies coverage of that issue, but coverage does not imply majority agreement.

\begin{definition}[Coverage]
A set of representatives $D$ covers an issue $s^i$ if $\exists d_l \in D$ such that $d_l^i = 1$.
\end{definition}

%We include for comparison Direct Democracy with incomplete participation.

% We assume that $|V|$ is odd to remove the need for tie-breaking under DD without any meaningful effect on our results.
% %
% Thus when $R$ is majority rule the outcome under Direct Democracy is $\vec{1}$.

%Don't introduce anything about the elections of weighting yet. Introduce it as it becomes relevant.

\section{Related Work}\label{sec:rr}

\citet{miller1969program}, inspired by \citet{tullock1967toward} and shareholder proxy voting, suggested an interactive democratic system for legislation that could take place at scale using computers. Miller lamented the lack of \emph{flexibility} in traditional Representative Democracy and sought to remedy this using a dynamic system of proxies, though admitting this was not conducive to creating legislation. Soon after,~\citet{Shubik1970aa} warned that electronic systems may accelerate the legislative process in undesirable ways and suggested holding every referendum twice to guarantee time for sufficient public deliberation. Our use of a fixed, elected set of representatives answers Miller's question of how to produce legislation, and rather than holding redundant referenda we propose to give the voters sufficient time to continue deliberation and alter their delegations after the representatives publicly vote.

Just before the dawn of the internet, \citet{tullock1992computerizing} revisited these ideas in a proposal that motivates the default distribution and delegation mechanism in FRD. The notion of the default distribution is also similar to the electoral weighting scheme proposed by~\citet{alger06proxy}, in which the weights of representatives are based on the preferences of voters expressed in the election, but these weights are fixed during the representatives' term. By contrast, in FRD the weight of each representative on each issue is not strictly determined by the election. We note that~\citet{DBLP:conf/atal/CohensiusMMMO17} took an analytical approach to studying a Proxy Voting model very close to that of~\citet{alger06proxy} for decision making with no election, infinite voters, spatial preferences, and assuming agents lie in a metric space.

The hallmark of an interactive democracy is that rather than adjudicating whether a direct or representative system is better for expressing the will of the voters and asserting it by fiat, the extent to which the system is direct or representative is itself a function of the will of the voters. Currently, the most well-known and well-studied form of Interactive Democracy is Liquid Democracy, which has been studied from an algorithmic perspective as a decision-making process in the AI and COMSOC literature~\citep{brill2018pairwise,kahng2021liquid,bloembergen2019rational,ChristoffG17,escoffier2019convergence,colley2021smart,becker2021can, markakis2021approval,colley2021multi, colley2022unravelling} and elsewhere~\citep{green2015direct,Ford02a,harding2022proxy,blum2016liquid,brill2018interactive,hardt2015google,harding2019incorporating,gersbach2022risky}. Unlike Liquid Democracy, FRD does not allow transitive delegations nor delegations to another voter, thereby violating the second axiom proposed by~\citet{green2015direct}. However, the notion of voluntary representatives can be maintained if desired for a particular application. Fractional delegations in FRD serve a similar function to that of the virtual committees proposed by~\citet{green2015direct}, although, in theory, FRD could incorporate virtual committees as well as many other mechanisms for delegating voting power.

The design of FRD is also largely based on work in probabilistic voting, binary aggregation, statistical decision theory, and computational social choice. In particular, work on the optimal weighting of experts~\citep{baharad2012beyond,nitzan2017collective,grofman1983determining,nitzan1982optimal,ben1997optimal}, the Condorcet Jury Theorem~\citep{grofman1983thirteen}, 
variable electorates~\citep{feld1984accuracy,smith1973aggregation,paroush1989robustness}, and optimal committee sizes~\citep{auriol2012optimal,karotkin2003optimum,magdon2018mathematical}. In FRD, one can view the voter delegations as a pseudo-tie breaking mechanism for the representatives or, conversely, see the default distribution as a way to dampen the variance in the outcome which occurs in Direct Democracy when the sample of participating voters is small or biased. Another view is that electing representatives is analogous to a compression algorithm~\citep{rodriguez2004societal}, which is the algorithmic version of John Adams's alleged intuition that the representatives should be a microcosm of the population (taken from~\citep{alger06proxy}). In this view, the delegations in FRD are a decompression mechanism where a higher delegation rate reduces the ``loss'' of representation. 

The recent works most similar to ours are that of \citet{pivato2020weighted}, \citet{soh2020approval}, and \citet{meir2021representative}, which each examine weighted representative voting systems in terms of their agreement with the voter majority. However, each considers weightings of representatives that are fixed across a set of binary issues.
Our evaluations are also similar to those of~\citep{Skow15a}, however, in their approval model the quality of the committee is measured as the sum of the voter proportion being represented for each issue, while we focus only on the total number of issue outcomes in alignment with the voter majority.

\section{Difficulties of Representation}\label{sec:RD}

In a Representative Democracy, the outcomes preferred by the voter majority are not known. However, even if it were known, electing a good set of representatives would still be hard.

\subsection{Complexity with Full Information}\label{sec:RD_complexity}
With a set of candidates $C$ with binary preferences over the same issues, it is NP-Hard to compute the subset of $k$ candidates that maximizes majority agreement.
To make matters worse, even maximizing coverage is NP-Hard.
We refer to the problems of electing $k$ candidates to maximize coverage and majority agreement when the preference of the voter majority is known for each issue as \emph{Max $k$-Coverage} and \emph{Max $k$-Majority Agreement}, respectively.

We begin by considering Max $k$-Coverage, and will use the NP-hardness of Max $k$-Coverage to prove the NP-hardness of Max $k$-Majority Agreement. In later sections, when we consider mechanisms where voters can re-weigh the representatives after the election, coverage of an issue will become the minimum necessary condition for there to exist a weighting that leads the outcome to agree with the voter majority.

\begin{problem}[Max $k$-Coverage]
Let $S = \{s^1, \ldots, s^r\}$ be a set of binary issues and $C = \{c_1, \ldots, c_m\}$ a set of candidates where candidate $c_l$ has preference $c_l^i \in \{0,1\}$ on issue $s^i$. The problem of Max $k$-Coverage is the problem of computing a subset of $k \leq m$ representatives $D \subseteq C$ that maximizes the number of covered issues, where issue $s^i \in S$ is covered if $\sum_{d_l \in D} d_l^i > 0$.
\end{problem}

\begin{theorem}\label{thm:maxkcover}
Max $k$-Coverage is NP-Hard.
\end{theorem}

\begin{proof}
Our proof of the hardness of Max $k$-Coverage is a straightforward Karp reduction via the NP-Hard problem of \textsc{MAX K-COVER}~\citep{feige1998threshold}.
The inputs to \textsc{MAX K-COVER} are a set $S = \{s^1, \ldots, s^r\}$ of $r$ points, a collection $C = \{c_1, \ldots, c_m\}$ of subsets of $S$, and an integer $k$. The objective of \textsc{MAX K-COVER} is to select $k$ subsets from $C$ such that their union has maximum cardinality.
Given an instance $(S, C, k)$ of \textsc{MAX K-COVER}, we create an instance $(\tilde{S}, \tilde{C}, \tilde{P}_C, k)$ of Max $k$-Coverage as follows.

For every point $s^i \in S$ create an issue $\tilde{s}^i \in \tilde{S}$, and for every subset $c_l \in C$ create a candidate $\tilde{c}_l \in \tilde{C}$. For all points $s^i \in S$ and subsets $c_l \in C$, if $s^i \in c_l$ then let $\tilde{c}_l^i = 1$, otherwise let $\tilde{c}_l^i = 0$. Let $k$ be the number of representatives to be elected. There is a one-to-one correspondence between the number of issues covered by out $k$ representatives and the cardinality of the corresponding subsets in the original \textsc{MAX K-COVER} instance. Therefore, any set of $k$ candidates that maximizes coverage corresponds exactly to a collection of $k$ subsets in our \textsc{MAX K-COVER} instance whose union has maximum cardinality.
\qed
\end{proof}

\begin{problem}[Max $k$-Majority Agreement]
Let $S = \{s^1, \ldots, s^r\}$ be a set of binary issues and $C = \{c_1, \ldots, c_m\}$ a set of candidates where candidate $c_l$ has preference $c_l^i \in \{0,1\}$ on issue $s^i$. The problem of Max $k$-Majority Agreement is the problem of computing a subset of $k \leq m$ representatives $D \subseteq C$, where $k$ is odd, that maximizes the number of issues on which the majority of representatives prefers $1$, i.e.
$$D = \argmax\limits_{C' \subseteq C, |C'| = k}|\{s^i \in S : \sum\limits_{c_l \in C'} c_l^i > \frac{|D|}{2}\}|$$
\end{problem}

\begin{theorem}
Max $k$-Majority Agreement is NP-Hard.
\end{theorem}

\begin{proof}
Our proof of the hardness of Max $k$-Majority Agreement is a Karp reduction via the problem of Max $k$-Coverage we proved is NP-Hard in Theorem \ref{thm:maxkcover}.
We will take an instance of Max $k$-Coverage and add $r+1$ additional issues to the original $r$ issues. In addition, we augment the candidate set with $k+1$ additional candidates in such a way that any subset of $2k+1$ candidates that maximizes agreement must contain the $k+1$ added candidates. The remaining $k$ winning candidates will correspond to the winning candidates for the original Max $k$-Coverage instance.

Suppose we have an instance of Max $k$-Coverage $(S, C, P_C, k)$ with $r = |S|$ issues and $m = |C|$ candidates.
We construct an instance of Max $k$-Majority Agreement $(\tilde{S}, \tilde{C}, \tilde{P}_C, \tilde{|D|})$ as follows. Let $\tilde{S} = (\tilde{s}^1, \ldots, \tilde{s}^{2r+1})$ be a set of $2r+1$ binary issues. Let $\tilde{C} = (\tilde{c}_1, \ldots, \tilde{c}_{m+k+1})$ be a set of $m+k+1$ candidates.
$C$ is made up of three types of candidates based on how we construct their preferences $\tilde{C} = (\tilde{c}_1, \ldots, \tilde{c}_m) \cup (\tilde{c}_{m+1}, \ldots, \tilde{c}_{m+k}) \cup (\tilde{c}_{m+k+1})$. The first $m$ candidates have the same preferences as the original $m$ candidates on the first $r$ issues and prefer $0$ for the added $2r+1$ issues. That is, $\forall l \leq m$, $\forall i \leq r$, $\tilde{c}_l^i = c_l^i$ and for all $i > r$, $\tilde{c}_l^i = 0$.
The next $k$ candidates unanimous prefer 1 on all issues: $\forall m < l \leq m+k$, $\forall i \leq 2r+1$, $\tilde{c}_l^i = 1$.
The final candidate, $\tilde{c}_{m+k+1}$, prefers $0$ on the first $r$ issues and $1$ on the remaining $r+1$ issues.
Lastly, let $\tilde{|D|} = 2k+1$.

Observe that any set $\tilde{D} \subseteq \tilde{C}$ of $2k+1$ candidates that maximizes majority agreement must contain candidates $(\tilde{c}_{m+1}, \ldots, \tilde{c}_{m+k})$ and $\tilde{c}_{m+k+1}$.
The majority of $\tilde{D}$ prefers $1$ on issues $r+1$ through $2r+1$ if and only if it contains $(\tilde{c}_{m+1}, \ldots, \tilde{c}_{m+k})$ and $\tilde{c}_{m+k+1}$. If even one of these $k+1$ candidates is not included in $\tilde{D}$, then the majority of $\tilde{D}$ will not prefer $1$ on any of these $r+1$ issues.
Since this constitutes more than half of the total issues, all of those candidates must be included in $\tilde{D}$.

The question remains how to select an additional $k$ candidates such that $\tilde{D}$ maximizes majority agreement on the first $r$ issues. The $k$ candidates $(\tilde{c}_{m+1}, \ldots, \tilde{c}_{m+k})$ prefer $1$ on the first $r$ issues but $\tilde{c}_{m+k+1}$ prefers 0, so to create a majority within $\tilde{D}$ on any of those issues requires one or more of the other $k$ winning candidates to prefer 1.
In other words, it is exactly the problem of using $k$ out of the original $m$ candidates to cover as many of the original $r$ issues as possible. Therefore, our solution to Max $k$-Majority Agreement on our constructed instance gives us the solution to the original Max $k$-Coverage problem.
\qed
\end{proof}

\subsection{Majority Agreement from Elections}\label{sec:RD_simulation}
We have shown that even when the preferences of the voter majority are known for every issue, it is computationally hard to select the optimal set of representatives from a set of candidates, in the general sense.
While solving Max $k$-Majority Agreement may be tractable for small enough instances, there are additional hurdles for electing representatives.
We now turn our attention to the partial-information setting where voters' preferences over issues are private. The voters report preferences over the candidates, and these electoral preferences are used to elect representatives. The preference of the voter majority is not known for any issue, and so we focus on how well different election methods approximate maximum majority agreement.

% \paragraph{Induced Electoral Preferences}
We give the election methods the best chances we can by assuming that voters' preferences over candidates are derived exactly from their agreement on the issues.
For brevity, we assume that the sets of voters and candidates are disjoint.
Voters will express their electoral preferences $P_{VC}$ as either approvals or complete rankings over the candidates.
When voters report approvals over the candidates, we assume that $v$ approves of $c$ if and only if they agree on more than half of the issues, i.e. $L(\vec{v}, \vec{c}) > \frac{1}{2}$.
When voters provide rankings of the candidates, we assume that $v$ prefers $c$ to $\hat{c}$ if they agree on a greater number of issues, i.e. $L(\vec{v},\vec{c}) > L(\vec{v},\hat{\vec{c}})$. If $L(\vec{v},\vec{c}) = L(\vec{v},\hat{\vec{c}})$, then $v$ breaks the tie privately (e.g. randomly).

Unfortunately, regardless of whether agents report approvals or rankings, no Condorcet-consistent election rule can approximate majority agreement.

\begin{theorem} \label{thm:anscombe}
No Condorcet-consistent election rule for selecting $k$ candidates in which voters report complete rankings over the candidates induced by their preferences over issues can provide a bounded approximation of majority agreement or coverage.
\end{theorem}

\begin{proof}
The example that proves Theorem \ref{thm:anscombe} has 11 voters, 11 issues, 2 candidates, and $k=1$. The instance in Example \ref{ex:anscombe} is derived from an example in~\citep{anscombe1976frustration}. We have two candidates; the ideal candidate $\vec{c_1} = \vec{1}$ and the worst conceivable candidate $\vec{c_2} = \vec{0}$. In this case, the majority of the voters would elect the worst conceivable candidate over the ideal candidate, and any Condorcet-consistent election rule must select the candidate preferred by the majority. Note also that the worst candidate receives a greater number of approval votes. This pathology arises because the majority of voters are in the minority on the majority of issues. Thus, if we only have duplicates of these two candidates for any $k$, there are cases in which the voters will elect representatives who achieve a majority agreement of 0 when a set of candidates exists who would achieve and agreement of 1. Each row represents the preference vector of a voter or candidate, with a column for each issue.

\begin{example}\label{ex:anscombe}
\resizebox{0.85\linewidth}{!}{
\begin{tabular}{llllllllllll}
         & $s^1$ & $s^2$ & $s^3$ & $s^4$ & $s^5$ & $s^6$ & $s^7$ & $s^8$ & $s^9$ & $s^{10}$ & $s^{11}$ \\
$v_1$    & 0     & 0     & 0     & 0     & 0     & 0     & 0     & 1     & 1     & 1        & 1        \\
$v_2$    & 1     & 1     & 1     & 1     & 0     & 0     & 0     & 0     & 0     & 0        & 0        \\
$v_3$    & 1     & 0     & 0     & 0     & 0     & 0     & 0     & 0     & 1     & 1        & 1        \\
$v_4$    & 1     & 1     & 0     & 0     & 0     & 0     & 0     & 0     & 0     & 1        & 1        \\
$v_5$    & 1     & 1     & 1     & 0     & 1     & 0     & 0     & 0     & 0     & 0        & 0        \\
$v_6$    & 0     & 0     & 0     & 1     & 1     & 1     & 1     & 0     & 0     & 0        & 0        \\
$v_7$    & 0     & 0     & 0     & 1     & 1     & 1     & 1     & 1     & 0     & 0        & 0        \\
$v_8$    & 1     & 1     & 1     & 0     & 0     & 1     & 1     & 1     & 1     & 1        & 1        \\
$v_9$    & 1     & 0     & 1     & 1     & 1     & 1     & 1     & 1     & 1     & 1        & 0        \\
$v_{10}$ & 0     & 1     & 1     & 1     & 1     & 1     & 1     & 1     & 1     & 0        & 1        \\
$v_{11}$ & 0     & 1     & 1     & 1     & 1     & 1     & 1     & 1     & 1     & 1        & 1        \\
$c_{1}$  & 1     & 1     & 1     & 1     & 1     & 1     & 1     & 1     & 1     & 1        & 1        \\
$c_{2}$  & 0     & 0     & 0     & 0     & 0     & 0     & 0     & 0     & 0     & 0        & 0       
\end{tabular}
}
\end{example}
\qed
\end{proof}

%While this example is deeply pathological, not all election rules are Condorcet-consistent, and it is only an edge case. We want to see how well election rules perform in expectation.

This pathological example can be generalized by duplicating voters and candidates. However, not all election rules are Condorcet-consistent. We proceed with an empirical investigation of the expected majority agreement from a number of election rules by simulation.

\paragraph{Simulated Elections}
We investigate majority agreement as a function of the numbers of voters, issues, candidates, issues, and representatives, as well as the choice of election rule. 
In all of our simulations, for all issues, the preference of every voter and candidate is $1$ with probability $\frac{1}{2}$. In other words, the preferences of every agent on every issue are determined by the flip of a fair coin. Thus, if the representatives were selected randomly from the candidates (i.e. sortition), that the expected majority agreement would be $0.5$. This assumption isolates the effects of the election, since any observed majority agreement beyond random chance is due entirely to the election process.
Our assumption of uniformly random voter preferences also implies that the size of the voter majority relative to the voter minority on an issue tends to be small. Let $V_1$ be the voters who vote for $1$ on an issue and $V_0$ be those who vote for 0. If we think of the voters as voting in some arbitrary sequence then the value of $|V_1|-|V_0|$ is modeled by an unbiased random walk, and so $\mathbb{E}[(|V_1|-|V_0|)^2] = |V|$.

As before, we assume that voters' preferences over candidates are induced by their agreement on the issues. Our assumption that the electoral preferences are derived without uncertainty is a best-case assumption for the election rules.
In additional to approvals and rankings, we now add the option for voters to weight the candidates using normalized weights. These weights provide strictly greater information than rankings.
With weights, $v_j$ assigns each candidate $c_l$ a weight of $w_{jl} = \frac{L(\vec{v}_j, \vec{c}_l)}{\sum_{c' \in C} L(\vec{v}_j, \vec{c}')}$. The weight attributed to each candidate is the sum of weights assigned to them, and the $k$ candidates with the greatest total weight are elected, with ties broken lexicographically.

\begin{figure*}[ht!]
\centering
\begin{subfigure}{0.33\linewidth}
	\centering
	\includegraphics[width=\linewidth]{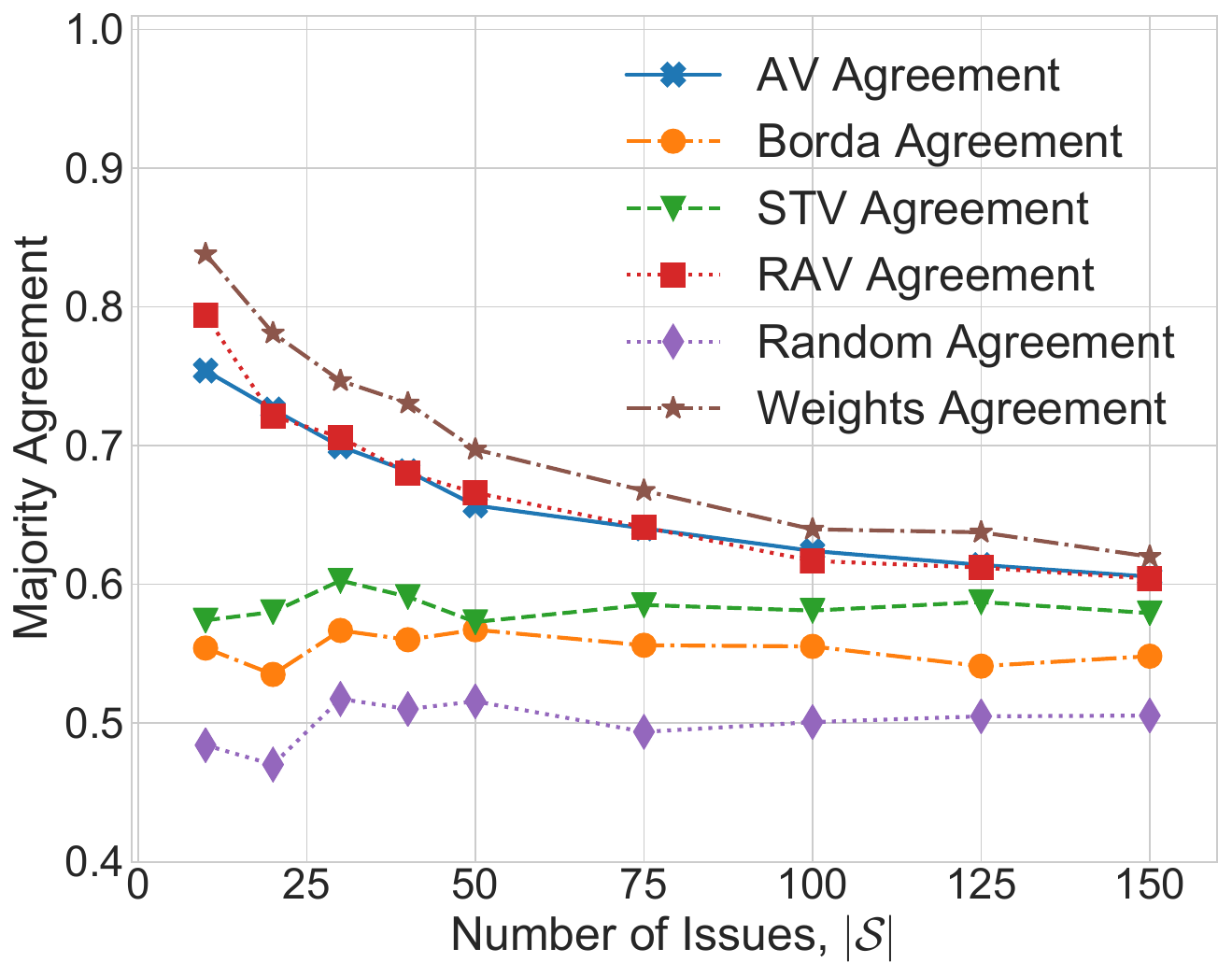}
	\caption{{\tiny Varying number of issues with $k=21, |C|=60, |V| = 501$.}}
	\label{fig:vary_issues}
\end{subfigure}
\hfill
\begin{subfigure}{0.33\linewidth} 
	\centering
	\includegraphics[width=\linewidth]{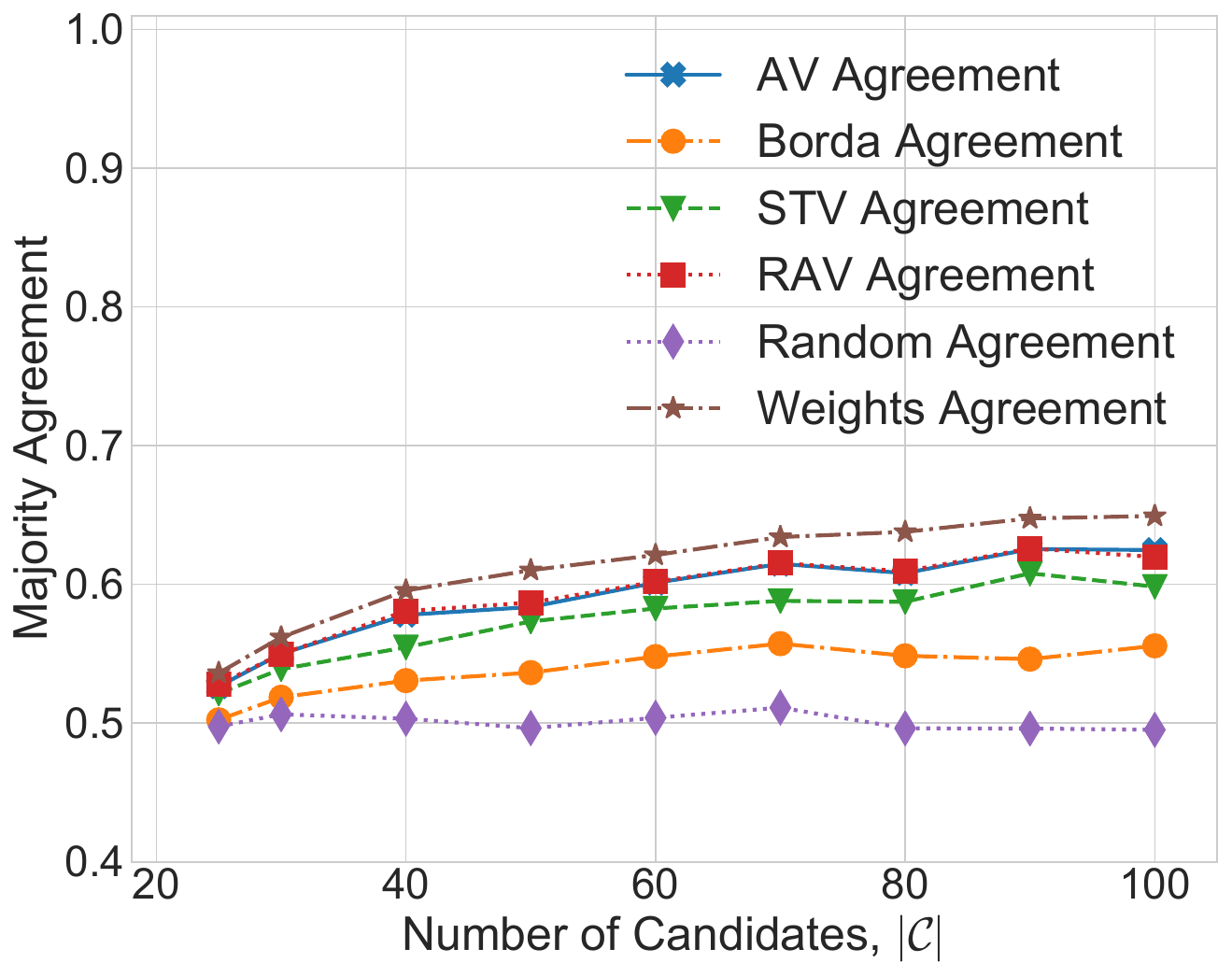}
	\caption{{\tiny Varying number of candidates with $k=21, |S|=150, |V| = 501$.}}
	\label{fig:vary_cands}
\end{subfigure}
\hfill
\begin{subfigure}{0.33\linewidth}
	\centering
	\includegraphics[width=\linewidth]{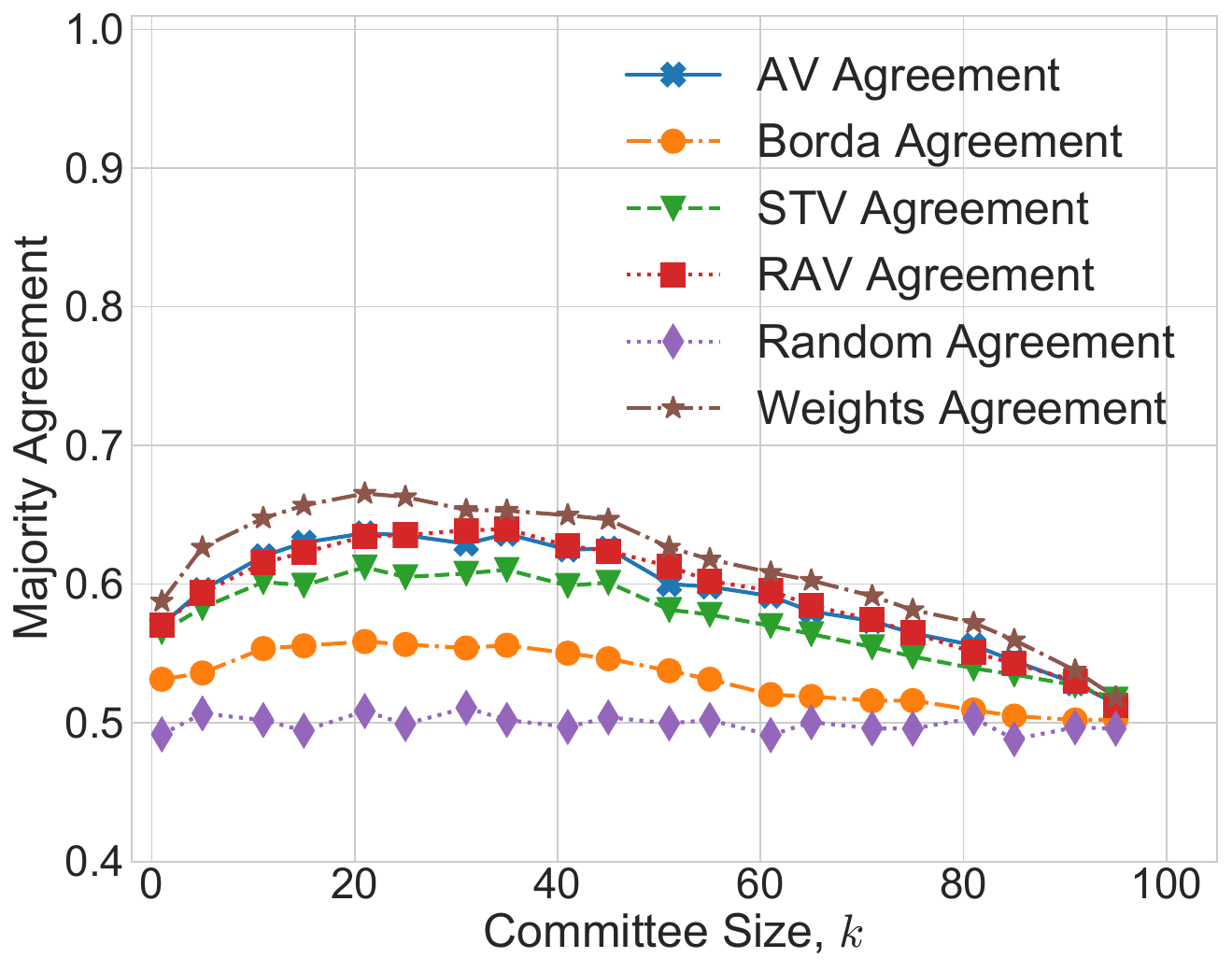}
	\caption{{\tiny Varying number of representatives $|C|=100, |S|=150, |V| = 501$.}}
	\label{fig:vary_k}
\end{subfigure}
\caption{Majority agreement as a function (a) the number of issues, (b) the number of candidates, and (c) the number of representatives.}
\vspace{-1.5em}
\label{fig:agree}
\end{figure*}

Figure \ref{fig:vary_issues}, Figure \ref{fig:vary_cands}, and Figure \ref{fig:vary_k} examine the numbers of issues, candidates, and representatives as independent variables, respectively. For each plot, the number of voters $|V|$ is held fixed at 501 because we did not observe significant dependence of majority agreement on the number of voters. Formal definitions of AV, RAV, STV, and Borda can be found in the book chapter by~\citep{zwicker2015introduction}.% and definitions of Chamberlain-Courant and $k$-Median can be found in~\citep{Skow15a}.

Turning first to Figure \ref{fig:vary_issues} we hold $|C| = 60$, $|k| = 21$ and vary $|S| \in \{15, \ldots, 150\}$ in steps of 15.  We see that for a small number of issues the AV, RAV, and the weighted voting rule can be expected to select a committee in agreement with the majority nearly 80\% of the time. However, as we add issues to the docket, the voting rules seem to converge around 60\%. In Figure \ref{fig:vary_cands} we fix $|k|=21, |S|=150$ and vary the number of candidates between $|C| \in \{21, \ldots, 101\}$ in steps of 5.  We observe again that AV, RAV, and weighted voting are the best followed closely by STV.  As we increase the number of candidates it is possible for the system to more frequently recover the will of the majority but this number does not climb above 65\% across all treatments.  Finally, in Figure \ref{fig:vary_k} we hold $|C|=100, |S|=150$ and vary $|k| \in \{1, 5, 11, 15, \ldots, 91, 95\}$.

While both Max $k$-Coverage and Max $k$-Majority Agreement are NP-Hard, majority agreement is a stronger condition because a majority agreement of $x \in [0,1]$ implies coverage of at least $x$. From the experiments in \ref{fig:agree}, we get a sense of just how much stronger of a condition majority agreement is. In all of our runs, coverage was 1.0 for all election rules and all combinations of parameters.
With Flexible Representative Democracy, allowing voters to weight representatives on each issue to increase agreement works in part because of the \emph{relative} ease with which coverage can be achieved by election rules compared to majority agreement. As long as an issue is covered, there exists a weighting such that the weighted majority of representatives agrees with the voter majority, even if the majority of representatives does not.

% \begin{itemize}
    % \item To start, we consider weighting the representatives based on the election preferences.
    % \item Simulation with weighting reps based on election preferences goes here. Discussion of simulation: does it do much when the weighting is not issue-specific?
% \end{itemize}

\section{Adding Flexibility}\label{sec:FRD}
Weighting representatives has the potential to increase their agreement with the voter majority, as long as issues are covered. To take advantage of the fact that coverage will tend to be greater than majority agreement, we look at different ways of weighting the representatives.
Along these lines, we'll extend the notion of majority agreement to include weighted majority voting.

\begin{definition}[Weighted Majority Agreement]
The weighted majority agreement between a set of voters $V$ and a set of representatives $D$ with weighting $W$ is the fraction of issues on which the weighted majority of reps agrees with the voter majority, i.e. $L(R(P_V), R(P_D))$ where $R$ is the weighted majority rule with random tie-breaking.
\end{definition}

The most basic way to weight the reps would be based on the agents' preferences over the candidates during the election. Unfortunately, as we will show in Section \ref{sec:FRD_simulation}, weighting the representatives based on the election profile $(P_{VC})$ offers little hope under our preference model. Instead, we allow the voters to weight the representatives on an issue-by-issue basis, if they choose.

Each voter gets a single, divisible voting unit on every issue. They can divide it among the representatives as they please. The weight of a representative is the sum (or average) of voting units given to them by the voters. The distribution of a voter's token among the representatives is referred to as \emph{delegation}, in line with the literature on ``delegative voting." Of course, if on every issue every voter delegates their full voting unit to a representative who agrees with them, then a weighted majority vote among the representatives exactly emulates Direct Democracy.  But what happens if a voter does not use their option to delegate their vote token among the representatives on one or more issues?
In standard proxy voting, if a voter does not choose a proxy, then they abstain.
The distinguishing feature of Flexible Representative Democracy is that it permits a default assignment of voting units from voters to representatives from which the voters may deviate.
The default distribution can be the uniform distribution, or it can be based on election preferences. Since election preferences do not appear to offer any benefit under our preference model, we will focus on the uniform distribution as the default when issue-specific delegations are allowed.
Under the uniform default distribution, the default evenly splits every voter's token among the representatives on every issue. If no agents shift their vote tokens on any issue, we have exactly a Representative Democracy. Here, Flexible Representative Democracy interpolates directly between Representative Democracy and Direct Democracy at the will of the voters.
We look first at basic features of Flexible Representative Democracy in a deterministic setting before considering probabilistic participation by the voters.

\subsection{Deterministic Delegation}\label{sec:FRD_deterministic}
Consider a single issue $s^i$.
Let $V_1 = \{v_j \in V : v_j^i = 1\}$ be the set of voters in the voter majority, and $V_0 = V \backslash V_1$ be the minority. Similarly, let $D_1 = \{d_l \in D : d_l^i = 1\}$, and $D_0 = D \backslash D_1$.

We denote by $\lambda_1$ the number of voters in $V_1$ who use their option to delegate, and $\lambda_0$ the delegators in the minority.
We say that a voter's delegation is \emph{incisive} if the voter delegates their entire voting unit to representatives who agree with them on that issue.
%
% The first question we have to answer is how large must $\lambda_1$ be to guarantee agreement with only incisive delegations (when $\lambda_0 = |V_0|$).

If the default is abstention, then agreement is guaranteed if the number of delegators in the majority is greater than the number of delegators in the minority, just like traditional proxy voting. The minority can push the outcome in their favor by delegating at a higher rate than the majority.

The first question we ask is, with a uniform default weighting, how many representatives have to agree with the voter majority for the agreement to be guaranteed. That is, if all minority voters delegate and none of the majority voters do, when does the uniform default guarantee agreement?

The weighting of the representatives depends on whether the representatives also have voting units to delegate to themselves. For brevity, we will assume the representatives do not, so if a representative receives no voting units from the voters, they have a weight of zero.
For consistency, we assume the sets of voters and representatives are disjoint, just as we assume the sets of voters and candidates are disjoint in our simulations.

\begin{example}\label{ex}
Consider an FRD instance with two issues $S = \{s^1, s^2\}$, three voters $V = \{v_1, v_2, v_3\}$, and three representatives $D = \{d_1, d_2, d_3\}$. Below, the solid arrows from voter to representative indicate incisive delegations, and any voter without an arrow sticks with the default uniform distribution on that issue. The voter and representative preferences are given in the tables above and below the agents. We denote by $X_1^i$ the total voting units assigned to representatives who agree with the voter majority on issue $s^i$ and by $W_l^i$ the weight of representative $d_l$ on $s^i$.

\begin{tikzpicture}[scale=0.4]

%%%WORSENING EXAMPLE%%%		
%Horizontal lines for table
\draw (4,14) -- (10,14);
\draw (4,12) -- (10,12);
\draw (4,10) -- (10,10);

\draw (4,6) -- (10,6);
\draw (4,4) -- (10,4);
\draw (4,2) -- (10,2);
\draw (4,0) -- (10,0);

%Vertical lines for table
\draw (4,0) -- (4,6);
\draw (6,0) -- (6,6);
\draw (8,0) -- (8,6);
\draw (10,0) -- (10,6);

\draw (4,10) -- (4,14);
\draw (6,10) -- (6,14);
\draw (8,10) -- (8,14);
\draw (10,10) -- (10,14);

%Voter Preferences
\node (v1) at (5,11) {$v_1$};
\node (v2) at (7,11) {$v_2$};
\node (v3) at (9,11) {$v_3$};

\node at (5,13) {$1$};
\node at (7,13) {$1$};
\node at (9,13) {$0$};

%Representative Preferences
\node (d1) at (5,5) {$d_1$};
\node (d2) at (7,5) {$d_2$};
\node (d3) at (9,5) {$d_3$};

\node at (5,3) {$1$};
\node at (7,3) {$1$};
\node at (9,3) {$0$};

\node at (5,1) {$\nicefrac{2}{3}$};
\node at (7,1) {$\nicefrac{2}{3}$};
\node at (9,1) {$\nicefrac{5}{3}$};

%Row Labels
\node (vji) at (3,13) {$v_j^i$};
\node (vj) at (3,11) {$v_j$};

\node (dl) at (3,5) {$d_l$};
\node (dli) at (3,3) {$d_l^i$};
\node (wli) at (3,1) {$W_l^i$};

%Delegation
\draw [ultra thick, ->] (9,10) -> (9,6);

%Issue
\node (s1) at (7,15) {Issue $s^1$};

%Outcome
\node (X1) at (7,-1.5) {$X^1_1 = \nicefrac{4}{3} < \frac{|V|}{2}$};
% \node (y1) at (7,-3) {$\out^1 = 0$};

%%%IMPROVEMENT EXAMPLE%%%		
%Horizontal lines for table
\draw (13,14) -- (19,14);
\draw (13,12) -- (19,12);
\draw (13,10) -- (19,10);

\draw (13,6) -- (19,6);
\draw (13,4) -- (19,4);
\draw (13,2) -- (19,2);
\draw (13,0) -- (19,0);

%Vertical lines for table
\draw (13,0) -- (13,6);
\draw (15,0) -- (15,6);
\draw (17,0) -- (17,6);
\draw (19,0) -- (19,6);

\draw (13,10) -- (13,14);
\draw (15,10) -- (15,14);
\draw (17,10) -- (17,14);
\draw (19,10) -- (19,14);

%Voter Preferences
\node (v1) at (14,11) {$v_1$};
\node (v2) at (16,11) {$v_2$};
\node (v3) at (18,11) {$v_3$};

\node at (14,13) {$1$};
\node at (16,13) {$1$};
\node at (18,13) {$0$};

%Representative Preferences
\node (d1) at (14,5) {$d_1$};
\node (d2) at (16,5) {$d_2$};
\node (d3) at (18,5) {$d_3$};

\node at (14,3) {$1$};
\node at (16,3) {$0$};
\node at (18,3) {$0$};

\node at (14,1) {$\nicefrac{5}{3}$};
\node at (16,1) {$\nicefrac{2}{3}$};
\node at (18,1) {$\nicefrac{2}{3}$};

%Row Labels
%\node (vji) at (12,13) {$v_j^i$};
%\node (vj) at (12,11) {$v_j$};
%
%\node (cj) at (12,5) {$c_j$};
%\node (cji) at (12,3) {$c_j^i$};
%\node (wji) at (11.5,1) {$weight$};

%Delegation
\draw [ultra thick, ->] (14,10) -> (14,6);

%Issue
\node (s2) at (16,15) {Issue $s^2$};

%Outcome
\node (X2) at (16,-1.5) {$X^2_1 = \nicefrac{5}{3} > \frac{|V|}{2}$};
% \node (y2) at (16,-3) {$\out^2 = 1$};

\end{tikzpicture}

On issue $s^1$, the representative majority agrees with the voter majority, so majority voting would yield agreement. However, since only the voter in the minority delegates ($v^1_3$), the weighted majority of representatives now decides the outcome in favor of the voter minority ($X^1_1 < \frac{|V|}{2}$). This can occur if the number of voters in the minority is large enough, the number of representatives who agree with the voter minority is large enough, and the voters in the minority delegate at a substantially higher rate than the voters in the majority.

On issue $s^2$ the representative majority disagrees with the voter majority so the majority voting outcome (without delegations) would be, regrettably, $0$. Looking again at the figure we see the delegations flip the result to what would be achieved by Direct Democracy ($X^2_1 > \frac{|V|}{2}$). Hence, FRD can improve the outcomes over Representative Democracy as measured by agreement with Direct Democracy. Fortunately, for both $s^1$ or $s^2$, if any two, or all three, of the voters delegate incisively, the outcome will always agree with the voter majority.
\end{example}

\begin{proposition}\label{prop:cantovercome}
If the representatives are weighted uniformly by default, all voters in the minority delegate incisively $(\lambda_0 = |V_0|)$, and none of the majority voters delegate $(\lambda_1 = 0)$, agreement is guaranteed when the number of representatives who agree with the voter majority $(D_1)$ is greater than $\frac{|D|}{2} \cdot \frac{|V|}{|V_1|}$.
\end{proposition}

\begin{proof}
We want to know under what conditions $X_1 > X_0$.
$X_1 = |V_1| \cdot \frac{|D_1|}{|D|}$ and $X_0 = |V_1| \cdot \frac{|D_0|}{|D|} + |V_0|$.

\begin{align*}
    X_1 = |V_1| \cdot \frac{|D_1|}{|D|} &> |V_1| \cdot \frac{|D_0|}{|D|} + |V_0| = X_0\\
    |V_1| \cdot \frac{|D_1| - |D_0|}{|D|} &> |V|-|V_1|\\
    |V_1| \cdot \left(\frac{|D_1| - |D_0|}{|D|} + 1\right) &> |V|\\
    |V_1| \cdot \left(\frac{|D_1| - |D_0| + |D|}{|D|}\right) &> |V|\\
    |V_1| \cdot \frac{2|D_1|}{|D|} &> |V|\\
    |D_1| &> \frac{|D|}{2}\cdot \frac{|V|}{|V_1|}
\end{align*}
\qed
\end{proof}

Now suppose the condition in Proposition \ref{prop:cantovercome} is not satisfied. If $|D_1| > |D_0|$, then the minority voters must delegate at a sufficiently higher rate than the majority voters for the weighted majority to no longer agree with the voter majority. If $|D_0| > |D_1|$, the majority delegation rate must be sufficiently higher than that of the minority to guarantee agreement. We make this statement more precise in Proposition \ref{prop:highenough}.

\begin{proposition}\label{prop:highenough}
If the representatives are weighted uniformly by default, the weighted majority of representatives agrees with the voter majority whenever $\lambda_1  > \frac{2 \lambda_0 |D_1| - (|V_1| - |V_0|)(|D_1| - |D_0|)}{2|D_0|}$.
\end{proposition}

\begin{proof}
The total voting units assigned to representatives in $D_1$ and $D_0$ are $X_1$ and $X_0$, respectively.
% $X_1 = \lambda_1 + \frac{|D_1|}{|D|}(|V_1| - \lambda_1) + \frac{|D_1|}{|D|}(|V_0| - \lambda_0)$ and
% $X_0 = \lambda_0 + \frac{|D_0|}{|D|}(|V_1| - \lambda_1) + \frac{|D_0|}{|D|}(|V_0| - \lambda_0)$, respectively.

\begin{align*}
    X_1 &> X_0\\
    \lambda_1 + \frac{|D_1|}{|D|}(|V_1| - \lambda_1) + \frac{|D_1|}{|D|}(|V_0| - \lambda_0) &> \lambda_0 + \frac{|D_0|}{|D|}(|V_1| - \lambda_1) + \frac{|D_0|}{|D|}(|V_0| - \lambda_0)\\
    \lambda_1 + \frac{|D_1| - |D_0|}{|D|}(|V_1| - \lambda_1) &> \lambda_0 + \frac{|D_0| - |D_1|}{|D|}(|V_0| - \lambda_0)\\
    \lambda_1 |D| + (|D_1| - |D_0|)(|V_1| - \lambda_1) &> \lambda_0 |D| - (|D_1| - |D_0|)(|V_0| - \lambda_0)\\
    2 \lambda_1 (|D_0|) + |V_1|(|D_1| - |D_0|) &> 2 \lambda_0 |D_1| + |V_0|(|D_1| - |D_0|)\\
    2 \lambda_1 (|D_0|) &> 2 \lambda_0 |D_1| - (|V_1| - |V_0|)(|D_1| - |D_0|)\\
    \lambda_1  &> \frac{2 \lambda_0 |D_1| - (|V_1| - |V_0|)(|D_1| - |D_0|)}{2|D_0|}
\end{align*}
\qed
\end{proof}

\subsection{Probabilistic Delegation}\label{sec:FRD_probabilistic}
We now investigate what happens if each voter chooses to delegate with some fixed individual probability. These results gives us an idea of how motivated or attentive voters must be to improve the outcome of FRD over Representative Democracy.

As all issues are independent, we will consider a single issue. Suppose each voter $v_j \in V$ chooses to delegate incisively with independent probability $p_j$ and leaves their voting unit distributed uniformly with probability $1 - p_j$. Let $x_j \in [0,1]$ be the total voting units $v_j$ assigns to candidates in $D_1$. If $v_j$ does not delegate then $x_j = \frac{|D_1|}{|D|}$, if $v_j$ delegates incisively and is in the voter majority then $x_j = 1$, and if $v_j$ delegates incisively and is in the voter minority then $x_j = 0$. Let $X_1 = \sum\limits_{v_j \in V} x_j$ be the total voting units assigned to $D_1$ via both delegation and the uniform default. Let $\mu = E[X_1] = \sum_{v_j \in V_1} \left(p_j + (1 - p_j)\frac{|D_1|}{|D|}\right) + \sum_{v_j \in V_0} (1 - p_j)\frac{|D_1|}{|D|}$ be the expected value of the total voting units assigned to representatives who agree with the voter majority.

\begin{theorem}\label{thm:prob}
Suppose there is an odd number of voters $|V|$, odd number of representatives $k=|D|$, a uniform default, no abstentions, and only incisive delegations. Suppose further that each voter $v_j \in V$ delegates with probability $p_j$ on an issue such that $\mu > \nicefrac{|V|}{2}$. Then the probability that the outcome agrees with the voter majority is bounded by $P(\text{weighted majority agreement}) \geq 1 - e^{-(|V|-2\mu)^2/4|V|}$.
\end{theorem}

\noindent
\textit{Proof Sketch: }
Let $y \in \{0,1\}$ denote the outcome of the weighted majority vote by the representatives.
The probability that the outcome agrees with the voter majority is $P(y=1) = P(X_1 > |V|/2) + P(y = 1 | X_1 = |V|/2) \cdot P(X_1 = |V|/2)$ where $P(y = 1 | X_1 = |V|/2)$ is due to the tie-breaking mechanism. First we show that with odd voters, odd representatives and only incisive delegations there can be no ties. Namely, $X_1 \neq X_0 = |V| - X_1$. This proof is due to parity and holds regardless of the delegation rate. Without ties, we simply need to determine $P(X_1 > |V|/2)$. We use a Chernoff inequality to provide a lower bound on this value based on the delegation probabilities $p_j$ of all voters. 
%See Section \ref{sec:ext} for discussion about tie-breaking 

\begin{proof}
Recall that $x_j \in [0,1]$ is the amount of voting units voter $v_j$ assigns to representatives who agree with the voter majority on an issue and $X_1 = \sum_{v_j \in V} x_j$. Given some tie breaking rule, we have that $P(y=1) = P(X_1 > |V|/2) + P(y = 1 | X_1 = |V|/2) \cdot P(X_1 = |V|/2)$. First we show that $P(X_1 = |V|/2) = 0$, then we give a lower bound for $P(X_1 > |V|/2)$.

\begin{lemma} \label{lemma_parity}
If $|V|$ is odd, $|D|$ is odd, and all delegations are incisive, then no ties are possible.
\end{lemma}

\begin{proof}
Let $x'_j = |D| \cdot x_j$ where $x_j \in [0,1]$ is the voting units voter $v_j$ assigns to candidates who agree with the voter majority on an issue via default or delegation. 
If $v_j$ does not delegate then $x'_j = |D_1|$, if $v_j$ delegates incisively and is in the voter majority then $x'_j = |D|$, and if $v_j$ delegates incisively and is in the voter minority then $x'_j =0$. Therefore, $ \forall j: x'_j \in \{0, |D_1|, |D|\}$.
Let $X'_1 = \sum\limits_{v_j \in V} x'_j$ and $X'_0 = \sum\limits_{v_j \in V} (|D|-x'_j)$. Then $X'_1, X'_0$ are non-negative integers and $X'_1 + X'_0 = |D|\cdot|V|$. Since $|D|\cdot|V|$ is odd, it must be that $X'_1$ and $X'_0$ have opposite parity and so they cannot be equal. Therefore $X_1 = \frac{X'_1}{|D|} \neq X_0 = \frac{X'_0}{|D|}$, meaning the total amounts of weight delegated to the representatives on either side of the issue cannot be equal, so no ties may occur.
\qed
\end{proof}

Given that no ties are possible, we have that $P(y = 1) = P(X_1 > |V|/2)$. Remember that $X_1 = \sum\limits_{v_j \in V} x_j$ where $x_j$ is the total weight that $v_j$ delegates to representatives who agree with the voter majority. If $v_j^i = 1$ then $E[x_j] = p_j + (1 - p_j)\frac{|D_1|}{|D|}$, else if $v_j^i = 0$ then $E[x_j] = (1 - p_j)\frac{|D_1|}{|D|}$. Let $\mu = E[X_1]$ be the expected total weight assigned to representatives who agrees with the voter majority. 
$$\mu = \sum\limits_{v_j \in V_1} \left(p_j + (1 - p_j)\frac{|D_1|}{|D|}\right) + \sum\limits_{v_j \in V_0} (1 - p_j)\frac{|D_1|}{|D|}$$

We now use the fact that $P(X_1 > |V|/2) = 1 - P(X_1 \leq |V|/2)$. Let $\delta = (2\mu - |V|)/2\mu$. If $\mu > |V|/2$, then $\delta > 0$. This allows us to apply a Chernoff bound to derive our lower bound:
\begin{align*}
    P(X_1 > |V|/2) &= 1 - P(X_1 \leq |V|/2)\\
    &= 1 - P(X_1 \leq (1-\delta)\mu) \\
    &\geq 1 - e^{-\delta^2\mu^2/|V|}\\
    &= 1 - e^{-(2\mu-N)^2/4|V|}
\end{align*}
\qed
\end{proof}

This bound depends on the condition that $\mu > |V|/2$. This condition requires that the delegation rate of the majority cannot be too small relative to the minority. Observe that this condition is satisfied when $\forall j: p_j = p$ and $|D_1| > \nicefrac{|D|}{2}$. Furthermore, as an increasing number of voters delegate incisively, we expect $\mu \rightarrow |V_1| > |V|/2$ regardless of $|D_1|$. Naturally, as the delegation rate increases, we observe our lower bound approach the ideal $1 - e^{-(2\mu-|V|)^2/4|V|} \rightarrow 1$. To find $\mu \leq |V|/2$, the voter majority cannot be too large compared to the voter minority, $|D_1|$ must be smaller than or somewhat close to $|D_0|$, and/or the voters in the majority must be significantly more apathetic towards delegation than voters in the minority. In other words, $\mu$ encodes the relative proportions used in Proposition \ref{prop:cantovercome} and Proposition \ref{prop:highenough} probabilistically.

\subsection{Simulated Delegation}\label{sec:FRD_simulation}
Finally, we provide an empirical comparison between Representative Democracy Flexible Representative Democracy with different issue-specific delegation behaviors by the voters.
We simulate instances where all voters delegate with equal probability. We refer to the total fraction of voters who delegate as the \emph{delegation rate} $(\alpha)$.

\begin{figure*}[ht!]
\begin{subfigure}{0.45\textwidth} 
\centering
\includegraphics[width=\linewidth]{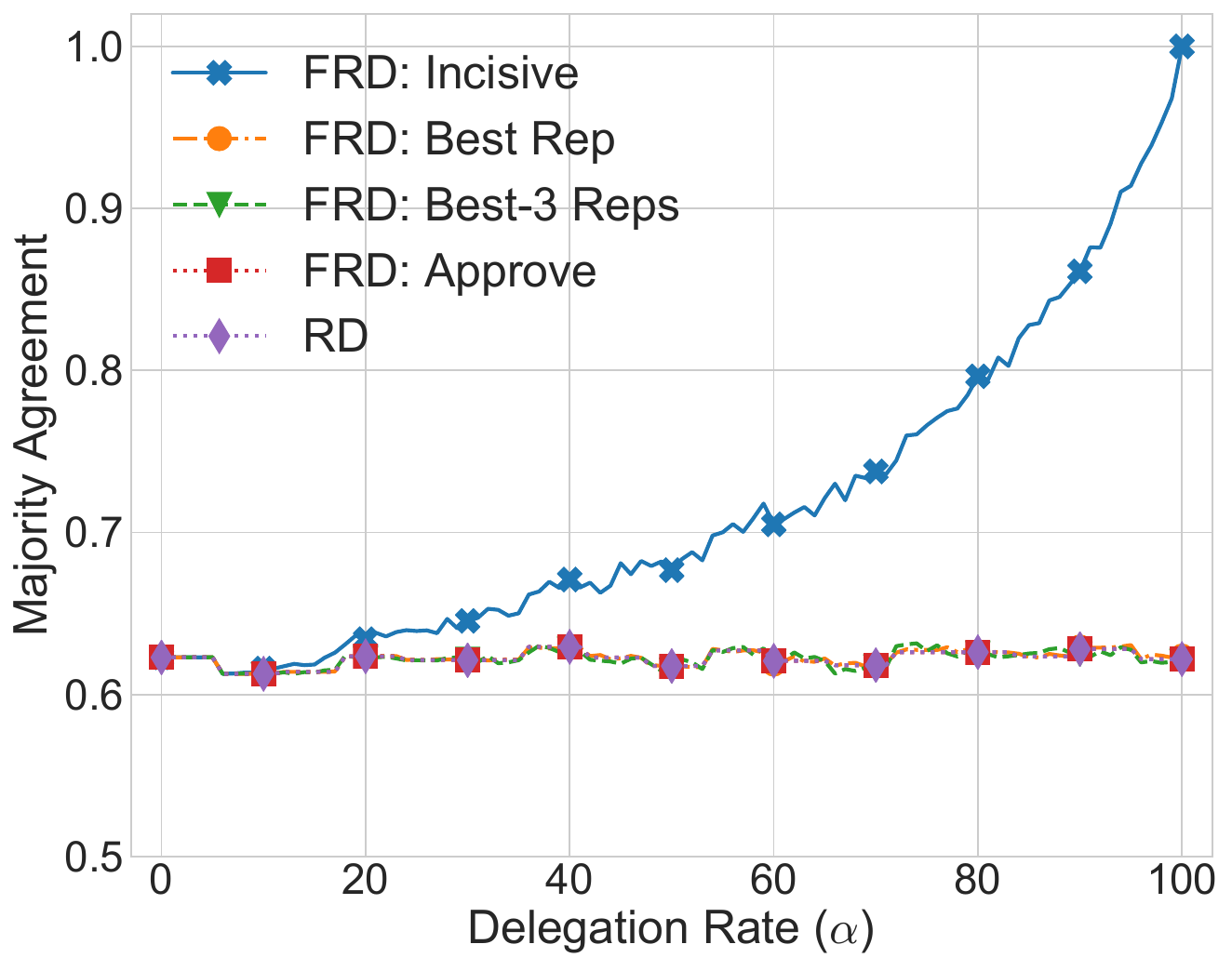}
\caption{Weighted Majority Agreement with Max Weight Election and Delegation}
\label{fig:delegation_rates}
\end{subfigure}
\hfill
\begin{subfigure}{0.45\textwidth} 
	\centering
	\includegraphics[width=\linewidth]{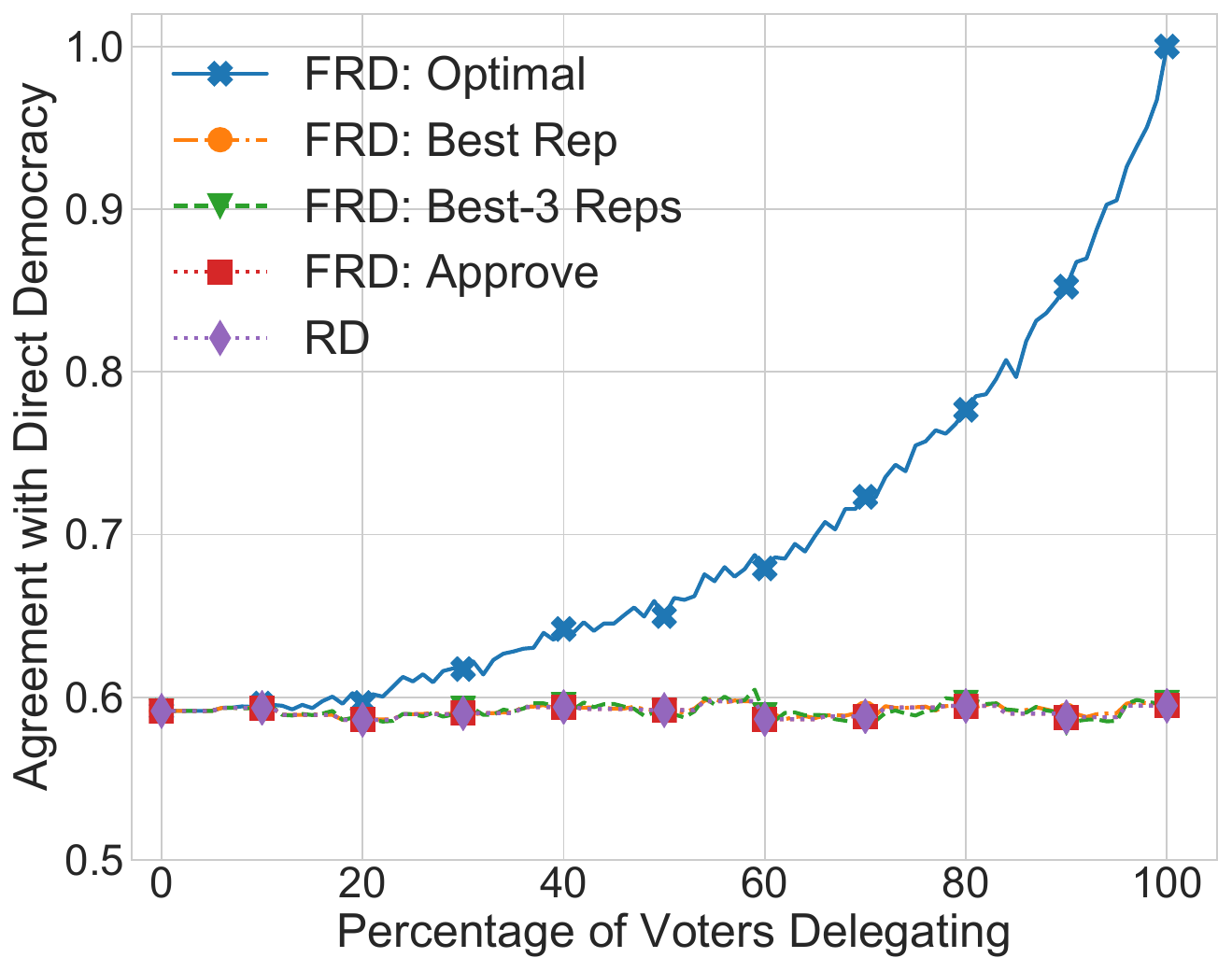}
	\caption{Weighted Majority Agreement with Borda Election and Delegation}
	\label{fig:borda_agreement}
\end{subfigure}
\end{figure*}

We use the same model to generate candidate and voter preferences as used in Section \ref{sec:RD_simulation}. For our simulated delegations we create instances with $|V|$ = 301, $|C|$ = 60, $|S|$ = 150, and $k$ = 21. In Figure \ref{fig:delegation_rates} representatives are elected using the Max-Weight election rule, and in Figure \ref{fig:borda_agreement} they are elected using Borda. We vary $\alpha \in [0,1.0]$ in increments of $0.01$ and for each setting of $\alpha$ we run 50 iterations. We plot the means of weighted majority agreement in Figure \ref{fig:delegation_rates}. A value of $1.0$ means that the outcomes of all issues are the same as they would be in a Direct Democracy with full participation. 

We compare Representative Democracy with four different delegation schemes: (1) \emph{Approve} where voters delegate evenly to the representatives whom they approved in the election across all issues; \emph{Best Rep} where voters delegate to their single most preferred representative (i.e. proxy) for all issues; \emph{Best-3 Rep} where voters delegate equally to their three most preferred representatives for all issues; and finally \emph{Incisive} where voters delegate incisively to a single representative with whom they agree on each individual issue. In all cases we use a uniform default weighting.

Most surprising is how little delegations that are not incisive and per-issue help emulate Direct Democracy. The \emph{Approve} delegation system is perhaps closest to the proposal of Proxy Voting espoused by~\citep{miller1969program} but does not improve weighted majority agreement in a meaningful way. Similarly interesting are the \emph{Best Rep} and \emph{Best-3 Rep} delegations, which also do not move the outcome towards the ideal of Direct Democracy.  Hence, we can see that issue-specific flexibility can be effectively used to improve agreement for voting systems in a way that other weighting schemes do not. Another striking result in Figure \ref{fig:delegation_rates} is how drastically FRD can improve agreement over Representative Democracy when voters are highly attentive. With as little as 60\% of the population delegating we can improve agreement by 10\% to a level not reached by any of the unweighted election rules in Figure \ref{fig:vary_issues}, Figure \ref{fig:vary_cands}, or Figure \ref{fig:vary_k} with 150 issues. When the delegation rate reaches 80\% we see an almost 20\% increase; eventually reaching 100\% when everyone delegates (since the issues are covered).

\section{Discussion}\label{sec:discusion}
While Direct Democracy with full participation may be held as an ideal, limitations on attention, information, and accessibility can make the use of proxies or elected representatives a more pragmatic alternative.
Unfortunately, handing all decision making power to a subset of the population is not without its downsides.
In particular, the collective choices of representatives may deviate from the majority views of the voters.
In a Representative Democracy, periodic elections and the ability to recall representatives during their term are meant to keep representatives accountable. But these methods are severely limited in their ability to keep representatives accountable on specific issues, and do little to promote civic engagement outside the election cycle.
Weighting the representatives, as with proxy voting, is similarly of limited benefit when the weights are not issue-specific.
Flexible Representative Democracy provides an alternative.

\paragraph{Voter Participation}
FRD does not require the direct participation of voters on every issue like Direct Democracy, but is able to benefit from their participation when possible. By contrast, Representative Democracy makes no direct use of the voters' preferences over issues, aside from the degree to which they are reflected in the election or influence the votes of representatives. While voters' preferences on an issue are not always accessible, they should not necessarily be discarded when they are available.
If we think of the election of representatives as a compression algorithm -- compressing the preference profile of the voters -- then FRD is a decompression algorithm.
When voter participation on a particular issue is low, the default distribution of voting units from the voters to the representatives -- as opposed to abstention by default -- prevents a small, potentially biased, sample of the voters from determining the outcome. Making the full withdrawal of voting units an active step rather than a passive one dampens the volatility that one might observe in the outcomes compared to proxy voting or Direct Democracy with only partial participation.
Moreover, voters are able to reflect their preference strengths to some degree by whether they put in the time, energy, and effort to delegate on various issues.

Unlike Representative Democracy, decision-making power is not handed over entirely to the representatives in FRD. The balance of power depends on how actively involved the voters choose to be in each decision, and how close the representatives are to being unanimous. The more divided the representatives are, the less voter participation is needed to determine the outcome. The closer the representatives are to unanimity, the more it takes for the voters to reverse the decision.

\paragraph{Public and Private Information}
When specifying an FRD mechanism it is necessary to note what information is public and what information is private.
One must be particularly concerned with what information voters have when delegating.
When the FRD mechanism uses a decision rule that weights the representatives as a subroutine as described in Section \ref{sec:prelims}, for a voter to know exactly the effect their delegation will have, they must know (1) the decision rule, (2) the current weights of the representatives, (3) how their delegation changes the weights of the representatives, and (4) the representatives' votes.
If delegation must stop before the representatives cast their ballots on an issue, then voters must base their delegations solely on the professed intentions of the representatives.
Permitting delegation after the representatives publicly cast immutable votes allows voters who take the opportunity to delegate incisively.
Unfortunately, for a voter to check the assignment of their individual voting units to representatives poses a privacy concern similar to the privacy concerns for voting receipts~\cite{sako1995receipt}.

There are several benefits to publicizing the weights of representatives. These weights provide information about what decisions are likely to be made and the relative support for different representatives. If multiple representatives vote similarly (e.g. come from the same political party) but have different weights, the weights can indicate factors other than preference similarity (e.g. perceived trustworthiness). Representatives have an incentive to inform the public to increase their weight to achieve their desired outcomes on issues.
If a representative does something unpopular, immoral, or illegal, a large decrease in their weight acts as a public censure without the need for an organized effort like a recall election. Moreover, if a representative's weight falls below a threshold on a number of issues they can be automatically recalled.

\paragraph{Direct Voting}
If voters are able to delegate after the representatives publicly cast their votes, why not allow voters to withdraw their voting units from the representatives entirely and cast a vote for themselves? This is certainly an option, but whether it is desirable depends on the application.
We do not provide a separate analysis of FRD with direct voting here because in our model with binary issues an incisive delegation is equivalent to a direct vote if the representatives are not unanimous.

If the representatives' votes tend to satisfy certain criteria that the voter profile as a whole does not, e.g. rationality axioms, it might be desirable to allow only delegations without the option to vote directly.
For example, suppose that the representatives generally vote on issues with more than two alternatives and cast their ballots in the form of rankings over the alternatives. If the voters' preferences as a whole tend not to satisfy single-peakedness but the representatives' vote profile does, then restricting the voters to delegation preserves single-peakedness. This is equivalent to allowing direct voting, but restricting the valid ballots that a voter can choose from to those selected by the representatives, or some linear combination of them.

\paragraph{Future Work}
The binary issues model and the analysis we provide are only the tip of the iceberg for studying Flexible Representative Democracy.
The most direct way to build on our results is to consider more sophisticated, and more realistic, models of preferences for binary issues and for election preferences. Assuming uniformly random preferences allowed us to isolate the effects of the election process and weighting schemes, but these types of preferences are rarely, if ever, observed empirically.
Assuming that voter preferences over candidates were derived precisely from their agreement was meant to give Representative Democracy the best chance of achieving high majority agreement, but this too should be relaxed.
Along the same lines, different participation rates, voter delegation behaviors, decision rules, and election rules remain to be examined.

Our core problem consisted of a set of independent, symmetric, binary issues to be decided. FRD mechanisms should be explored for asymmetric binary issues with a status quo, issues with more than two alternatives, and different representative ballot types. It also remains to be seen how FRD will perform when the outcomes of issues are logically dependent, as in the literature on binary aggregation. When issues are dependent one should expect voter preferences across issues to exhibit dependencies, and so preference models like conditional preference networks may be most appropriate.

For both binary issues and other issue types, objectives beyond agreement deserve to be examined.
Voter preferences may be assumed to derive from utilities in different ways, which can be used to measure the quality of outcomes.
On the other hand, FRD mechanisms can be evaluated in terms of their ability to uncover ground-truths, similar to some of the literature on Liquid Democracy and Condorcet Jury Theorems.
Our analysis did not examine strategic behavior by the voters, candidates, or representatives, but this possibility raises may interesting questions.

Lastly, we have not discussed implementing FRD mechanisms and the various privacy, security, and technological concerns that arise.

\subsubsection{Acknowledgements}
Nicholas Mattei was supported by NSF Awards IIS-RI-2007955, IIS-III-2107505, and IIS-RI-2134857, as well as an IBM Faculty Award and a Google Research Scholar Award. Ben Abramowitz was supported by the NSF under Grant \#2127309 to the Computing Research Association for the CIFellows Project.

\bibliographystyle{spbasic}
\bibliography{abb, abbshort, FRD_SCW.bib}

\begin{thebibliography}{57}
\providecommand{\natexlab}[1]{#1}
\providecommand{\url}[1]{{#1}}
\providecommand{\urlprefix}{URL }
\expandafter\ifx\csname urlstyle\endcsname\relax
  \providecommand{\doi}[1]{DOI~\discretionary{}{}{}#1}\else
  \providecommand{\doi}{DOI~\discretionary{}{}{}\begingroup
  \urlstyle{rm}\Url}\fi
\providecommand{\eprint}[2][]{\url{#2}}

\bibitem[{Alger(2006)}]{alger06proxy}
Alger D (2006) Voting by proxy. Public Choice 126(1/2):1--26,
  \urlprefix\url{http://www.jstor.org/stable/30026573}

\bibitem[{Anscombe(1976)}]{anscombe1976frustration}
Anscombe GE (1976) On frustration of the majority by fulfillment of the
  majority's will. Analysis 36(4):161--168

\bibitem[{Anshelevich et~al.(2021)Anshelevich, Fitzsimmons, Vaish, and
  Xia}]{anshelevich2021representative}
Anshelevich E, Fitzsimmons Z, Vaish R, Xia L (2021) Representative proxy
  voting. In: Proceedings of the AAAI Conference on Artificial Intelligence,
  vol~35, pp 5086--5093

\bibitem[{Auriol and Gary-Bobo(2012)}]{auriol2012optimal}
Auriol E, Gary-Bobo RJ (2012) On the optimal number of representatives. Public
  Choice 153(3-4):419--445

\bibitem[{Aziz et~al.(2017)Aziz, Brill, Conitzer, Elkind, Freeman, and
  Walsh}]{aziz2017justified}
Aziz H, Brill M, Conitzer V, Elkind E, Freeman R, Walsh T (2017) Justified
  representation in approval-based committee voting. Social Choice and Welfare
  48(2):461--485

\bibitem[{Baharad et~al.(2012)Baharad, Goldberger, Koppel, and
  Nitzan}]{baharad2012beyond}
Baharad E, Goldberger J, Koppel M, Nitzan S (2012) Beyond {C}ondorcet:
  {O}ptimal aggregation rules using voting records. Theory and decision
  72(1):113--130

\bibitem[{Becker et~al.(2021)Becker, D'Angelo, Delfaraz, and
  Gilbert}]{becker2021can}
Becker R, D'Angelo G, Delfaraz E, Gilbert H (2021) When can liquid democracy
  unveil the truth? arXiv preprint arXiv:210401828

\bibitem[{Behrens et~al.(2014)Behrens, Kistner, Nitsche, and
  Swierczek}]{behrens2014principles}
Behrens J, Kistner A, Nitsche A, Swierczek B (2014) The Principles of
  LiquidFeedback. Interacktive Demokratie

\bibitem[{Ben-Yashar and Nitzan(1997)}]{ben1997optimal}
Ben-Yashar RC, Nitzan SI (1997) The optimal decision rule for fixed-size
  committees in dichotomous choice situations: {T}he general result.
  International Economic Review pp 175--186

\bibitem[{Bloembergen et~al.(2019)Bloembergen, Grossi, and
  Lackner}]{bloembergen2019rational}
Bloembergen D, Grossi D, Lackner M (2019) On rational delegations in liquid
  democracy. In: Proceedings of the AAAI Conference on Artificial Intelligence,
  vol~33, pp 1796--1803

\bibitem[{Blum and Zuber(2016)}]{blum2016liquid}
Blum C, Zuber CI (2016) Liquid democracy: {P}otentials, problems, and
  perspectives. Journal of Political Philosophy 24(2):162--182

\bibitem[{Brandt et~al.(2016)Brandt, Conitzer, Endriss, Lang, and
  Procaccia}]{BCELP16a}
Brandt F, Conitzer V, Endriss U, Lang J, Procaccia AD (eds)  (2016) Handbook of
  Computational Social Choice. Cambridge University Press

\bibitem[{Brill(2018)}]{brill2018interactive}
Brill M (2018) Interactive democracy. In: Proc.~17th AAMAS, pp 1183--1187

\bibitem[{Brill and Talmon(2018)}]{brill2018pairwise}
Brill M, Talmon N (2018) Pairwise liquid democracy. In: Proc.~27th IJCAI, pp
  137--143

\bibitem[{Christoff and Grossi(2017)}]{ChristoffG17}
Christoff Z, Grossi D (2017) Binary voting with delegable proxy: An analysis of
  liquid democracy. In: Proc.~16th TARK, pp 134--150

\bibitem[{Cohensius et~al.(2017)Cohensius, Mannor, Meir, Meirom, and
  Orda}]{DBLP:conf/atal/CohensiusMMMO17}
Cohensius G, Mannor S, Meir R, Meirom EA, Orda A (2017) Proxy voting for better
  outcomes. In: Proc.~16th AAMAS, pp 858--866

\bibitem[{Colley(2021)}]{colley2021multi}
Colley R (2021) Multi-agent ranked delegations in voting. In: Proceedings of
  the 20th International Conference on Autonomous Agents and MultiAgent
  Systems, pp 1802--1804

\bibitem[{Colley et~al.(2021)Colley, Grandi, and Novaro}]{colley2021smart}
Colley R, Grandi U, Novaro A (2021) Smart voting. In: Twenty-Ninth
  International Joint Conference on Artificial Intelligence (IJCAI 2020),
  International Joint Conferences on Artifical Intelligence (IJCAI), pp
  1734--1740

\bibitem[{Colley et~al.(2022)Colley, Grandi, and
  Novaro}]{colley2022unravelling}
Colley R, Grandi U, Novaro A (2022) Unravelling multi-agent ranked delegations.
  Autonomous Agents and Multi-Agent Systems 36(1):1--35

\bibitem[{Domshlak et~al.(2011)Domshlak, H{\"u}llermeier, Kaci, and
  Prade}]{DHKP11a}
Domshlak C, H{\"u}llermeier E, Kaci S, Prade H (2011) Preferences in {AI}: An
  overview. AI 175(7):1037--1052

\bibitem[{Escoffier et~al.(2019)Escoffier, Gilbert, and
  Pass-Lanneau}]{escoffier2019convergence}
Escoffier B, Gilbert H, Pass-Lanneau A (2019) The convergence of iterative
  delegations in liquid democracy in a social network. In: International
  Symposium on Algorithmic Game Theory, Springer, pp 284--297

\bibitem[{Feige(1998)}]{feige1998threshold}
Feige U (1998) A threshold of ln n for approximating set cover. Journal of the
  ACM 45(4):634--652

\bibitem[{Feld and Grofman(1984)}]{feld1984accuracy}
Feld SL, Grofman B (1984) The accuracy of group majority decisions in groups
  with added members. Public Choice 42(3):273--285

\bibitem[{Ford(2002)}]{Ford02a}
Ford B (2002) Delegative democracy, unpublished Manuscript,
  http://brynosaurus.com/deleg/deleg.pdf

\bibitem[{F{\"u}rnkranz and H{\"u}llermeier(2010)}]{FuHu10a}
F{\"u}rnkranz J, H{\"u}llermeier E (2010) Preference Learning. Springer

\bibitem[{Gersbach et~al.(2022)Gersbach, Mamageishvili, and
  Schneider}]{gersbach2022risky}
Gersbach H, Mamageishvili A, Schneider M (2022) Risky vote delegation

\bibitem[{G{\"o}lz et~al.(2021)G{\"o}lz, Kahng, Mackenzie, and
  Procaccia}]{golz2021fluid}
G{\"o}lz P, Kahng A, Mackenzie S, Procaccia AD (2021) The fluid mechanics of
  liquid democracy. ACM Transactions on Economics and Computation 9(4):1--39

\bibitem[{Green-Armytage(2015)}]{green2015direct}
Green-Armytage J (2015) Direct voting and proxy voting. Constitutional
  Political Economy 26(2):190--220

\bibitem[{Grofman and Feld(1983)}]{grofman1983determining}
Grofman B, Feld SL (1983) Determining optimal weights for expert judgment. In:
  Information Pooling and Group Decision Making, pp 167--72

\bibitem[{Grofman et~al.(1983)Grofman, Owen, and Feld}]{grofman1983thirteen}
Grofman B, Owen G, Feld SL (1983) Thirteen theorems in search of the truth.
  Theory and Decision 15(3):261--278

\bibitem[{Harding(2019)}]{harding2019incorporating}
Harding J (2019) Incorporating preference information into formal models of
  transitive proxy voting

\bibitem[{Harding(2022)}]{harding2022proxy}
Harding J (2022) Proxy selection in transitive proxy voting. Social Choice and
  Welfare 58(1):69--99

\bibitem[{Hardt and Lopes(2015)}]{hardt2015google}
Hardt S, Lopes LC (2015) Google votes: {A} liquid democracy experiment on a
  corporate social network. Technical Disclosure Commons

\bibitem[{Kahng et~al.(2021)Kahng, Mackenzie, and Procaccia}]{kahng2021liquid}
Kahng A, Mackenzie S, Procaccia A (2021) Liquid democracy: An algorithmic
  perspective. Journal of Artificial Intelligence Research 70:1223--1252

\bibitem[{Karotkin and Paroush(2003)}]{karotkin2003optimum}
Karotkin D, Paroush J (2003) Optimum committee size: Quality-versus-quantity
  dilemma. Social Choice and Welfare 20(3):429--441

\bibitem[{KhudaBukhsh et~al.(2016)KhudaBukhsh, Xu, Hoos, and
  Leyton-Brown}]{khudabukhsh2009satenstein}
KhudaBukhsh AR, Xu L, Hoos HH, Leyton-Brown K (2016) {SATenstein}:
  {A}utomatically building local search sat solvers from components. AI
  232:20--42

\bibitem[{Magdon-Ismail and Xia(2018)}]{magdon2018mathematical}
Magdon-Ismail M, Xia L (2018) A mathematical model for optimal decisions in a
  representative democracy. Advances in Neural Information Processing Systems
  31

\bibitem[{Markakis and Papasotiropoulos(2021)}]{markakis2021approval}
Markakis E, Papasotiropoulos G (2021) An approval-based model for single-step
  liquid democracy. In: International Symposium on Algorithmic Game Theory,
  Springer, pp 360--375

\bibitem[{Meir et~al.(2021)Meir, Sandomirskiy, and
  Tennenholtz}]{meir2021representative}
Meir R, Sandomirskiy F, Tennenholtz M (2021) Representative committees of
  peers. Journal of Artificial Intelligence Research 71:401--429

\bibitem[{Miller(1969)}]{miller1969program}
Miller JC (1969) A program for direct and proxy voting in the legislative
  process. Public Choice 7(1):107--113

\bibitem[{Nitzan and Paroush(1982)}]{nitzan1982optimal}
Nitzan S, Paroush J (1982) Optimal decision rules in uncertain dichotomous
  choice situations. International Economic Review pp 289--297

\bibitem[{Nitzan and Paroush(2017)}]{nitzan2017collective}
Nitzan S, Paroush J (2017) Collective decision making and jury theorems. Oxford
  Handbook of Law and Economics pp 494--516

\bibitem[{Paroush and Karotkin(1989)}]{paroush1989robustness}
Paroush J, Karotkin D (1989) Robustness of optimal majority rules over teams
  with changing size. Social Choice and Welfare 6(2):127--138

\bibitem[{Pivato and Soh(2020)}]{pivato2020weighted}
Pivato M, Soh A (2020) Weighted representative democracy. Journal of
  Mathematical Economics 88:52--63

\bibitem[{Rodriguez et~al.(2004)Rodriguez, Steinbock
  et~al.}]{rodriguez2004societal}
Rodriguez MA, Steinbock DJ, et~al. (2004) Societal-scale decision making using
  social networks. In: North American Association for Computational Social and
  Organizational Science Conference Proceedings

\bibitem[{Sako and Kilian(1995)}]{sako1995receipt}
Sako K, Kilian J (1995) Receipt-free mix-type voting scheme. In: International
  Conference on the Theory and Applications of Cryptographic Techniques,
  Springer, pp 393--403

\bibitem[{Shubik(1970)}]{Shubik1970aa}
Shubik M (1970) On homo politicus and the instant referendum. Public Choice
  9:79--84

\bibitem[{Skowron(2015{\natexlab{a}})}]{Skow15a}
Skowron P (2015{\natexlab{a}}) What do we elect committees for? {A} voting
  committee model for multi-winner rules. In: Proc.~24th IJCAI, pp 1141--1147

\bibitem[{Skowron et~al.(2016)Skowron, Faliszewski, and Lang}]{SFL16a}
Skowron P, Faliszewski P, Lang J (2016) Finding a collective set of items:
  {F}rom proportional multi-representation to group recommendation. AI
  241:191--216

\bibitem[{Skowron(2015{\natexlab{b}})}]{skowron2015we}
Skowron PK (2015{\natexlab{b}}) What do we elect committees for? a voting
  committee model for multi-winner rules. In: Twenty-Fourth International Joint
  Conference on Artificial Intelligence

\bibitem[{Smith(1973)}]{smith1973aggregation}
Smith JH (1973) Aggregation of preferences with variable electorate.
  Econometrica: Journal of the Econometric Society pp 1027--1041

\bibitem[{Soh~Voutsa(2020)}]{soh2020approval}
Soh~Voutsa AC (2020) Approval voting \& majority judgement in weighted
  representative democracy. Available at SSRN

\bibitem[{Tullock(1967)}]{tullock1967toward}
Tullock G (1967) Toward a mathematics of politics. University of Michigan Press

\bibitem[{Tullock(1992)}]{tullock1992computerizing}
Tullock G (1992) Computerizing politics. Mathematical and Computer Modelling
  16(8-9):59--65

\bibitem[{Yu et~al.(2010)Yu, Werfel, and Nagpal}]{Yu:2010:CDM:1838186.1838192}
Yu CH, Werfel J, Nagpal R (2010) Collective decision-making in multi-agent
  systems by implicit leadership. In: Proc.~9th AAMAS, pp 1189--1196

\bibitem[{Zhu et~al.(2012)Zhu, Zhou, and Alkins}]{zhu2012group}
Zhu H, Zhou M, Alkins R (2012) Group role assignment via a {K}uhn--{M}unkres
  algorithm-based solution. IEEE Transactions on Systems, Man, and
  Cybernetics-Part A: Systems and Humans 42(3):739--750

\bibitem[{Zwicker(2015)}]{zwicker2015introduction}
Zwicker WS (2015) Introduction to voting theory. In: Brandt F, Conitzer V,
  Endriss U, Lang J, Procaccia AD (eds) Handbook of Computational Social
  Choice, Cambridge University Press, chap~2

\end{thebibliography}

\section{Appendix}\label{sec:appendix}

% \subsection{Approval Thresholds}\labe{sec:thresholds}

\subsection{Full Coverage}\label{sec:app:full_coverage}
In Section \ref{sec:RD_complexity}, we prove that Max $k$-Coverage is NP-Hard. However, for all voters to be able to delegate incisively, more than coverage is required. It is necessary that the representatives not be unanimous. In other words, there must be at least one representative who agrees with the voter majority and at least one representative who agrees with the voter minority on every issue. If we say that any issue on which the representatives are not unanimous is \emph{fully covered} then we want to know if Max $k$-Full Coverage is NP-Hard as well. We provide an affirmative proof below. 

\begin{problem}[Max $k$-Full Coverage]
Let $S = \{s^1, \ldots, s^r\}$ be a set of binary issues and $C = \{c_1, \ldots, c_m\}$ a set of candidates where candidate $c_l$ has preference $c_l^i \in \{0,1\}$ on issue $s^i$. The problem of Max $k$-Full Coverage is the problem of computing a subset of $k \leq m$ representatives $D \subseteq C$ that maximizes the number of fully covered issues, where issue $s^i \in S$ is fully covered if $0 < \sum_{d_l \in D} d_l^i < |D|$.
\end{problem}

\begin{theorem} \label{thm:full_coverage}
Max $k$-Full Coverage is NP-Hard.
\end{theorem}

\begin{proof}
We now prove the complexity of Max $k$-Full Coverage by polynomial-time reduction from Max $k$-Coverage. To do this we construct an instance of Max $k$-Full Coverage by adding an additional candidate $\hat{c}$, adding $r+1$ additional issues to the original $r$ issues, and desire a set of $k+1$ candidates. We show that in this new instance of Max $k$-Full Coverage the additional candidate must be selected in any optimal solution because they are uniquely required to cover the $r+1$ added issues, and the remaining $k$ candidates in the solution set correspond exactly to the optimal $k$ candidates in the solution to our original Max $k$-Coverage instance.

Given an instance $(S = \{s^1, \ldots, s^r\}, C = \{c_1, \ldots, c_m\}, k)$ of Max $k$-Coverage we construct an instance of Max $k$-Full Coverage as follows. Create a set of binary issues $\tilde{S}$ equal to $S$ augmented with $r+1$ additional binary issues so that $\tilde{S} = \{s^1, \ldots, s^r, s^{r+1} \ldots, s^{2r+1}\}$.
Create a set of candidates $\tilde{C} = C \cup \hat{c}$ where $\tilde{c}_l^i = c_l^i$ for all $c_l \in C$, for all $s^i \in \{s^1, \ldots, s^r\}$, and $\hat{c}^i = 0$ for all issues $s^i \in \{s^1, \ldots, s^r\}$
Let $\tilde{c}_l^i = 0$ for all $\tilde{c_l} \in \tilde{C} \backslash \{\hat{c}\}$ for issues $s^i \in \{s^{r+1}, \ldots, s^{2r+1}\}$ and let $\hat{c}^i = 1$ for all issues $s^i \in \{s^{r+1}, \ldots, s^{2r+1}\}$.
Our objective is to select a set $\tilde{D} \subseteq \tilde{C}$ of $k+1$ candidates from $\tilde{C}$ that maximizes full coverage. We will now prove that for all solutions $\tilde{D}$ to our new Max $k$-Full Coverage problem, $\tilde{D} = \{\hat{c}\} \cup D$ where $D$ is a set of $k$ candidates whose corresponding counterparts maximize coverage over issues $\{s^1, \ldots, s^r\}$ in our original Max $k$-Coverage instance.

\begin{lemma}
$\tilde{D}$ must contain $\hat{c}$
\end{lemma}

\begin{proof}
Clearly, the set $\{ \hat{c} \} \cup \{ \tilde{c}_l \}$ achieves full coverage for issues $\{s^{r+1}, \ldots, s^{2r+1}\}$ for any $\tilde{c}_l \in \tilde{C} \backslash \{\hat{c}\}$, and any set which does not contain $\hat{c}$ cannot fully cover (or cover) $\{s^{r+1}, \ldots, s^{2r+1}\}$. Since $\{s^{r+1}, \ldots, s^{2r+1}\}$ comprises more than half the issues, any set of $k+1$ candidates for $k \geq 1$ that maximizes the number of issues fully covered must contain $\hat{c}$.
\qed
\end{proof}

Given that $\hat{c}^i = 0$ for all issues $\{s^1, \ldots, s^r\}$, the set of candidates $D = \tilde{D} \backslash \{\hat{c}\}$, which maximizes full coverage for issues $\{s^1, \ldots, s^r\}$ is the set of $k$ candidates which maximizes coverage over issues in $\{s^1, \ldots, s^r\}$. Therefore, the $k$ candidates corresponding to $D$ are the solution to our original instance of Max $k$-Coverage and given the solution $D$ to Max $k$-Coverage we simply add $\hat{c}$ to find $\tilde{D}$.
\qed
\end{proof}

\subsection{NP-Hard Election Rules}\label{sec:app:np_hard_election}
The election rules examined in Section \ref{sec:RD_simulation} all have polynomial-time winner determination. One may wish to compare the performance of election rules with more complex winner determination. We provide a simulated comparison with the addition of the Chamberlin-Courant and $k$-Median election rules. Definitions of Chamberlin-Courant and $k$-Median can be found in \citep{Skow15a}.

\begin{figure}[h]
    \centering
    \includegraphics[width=\linewidth]{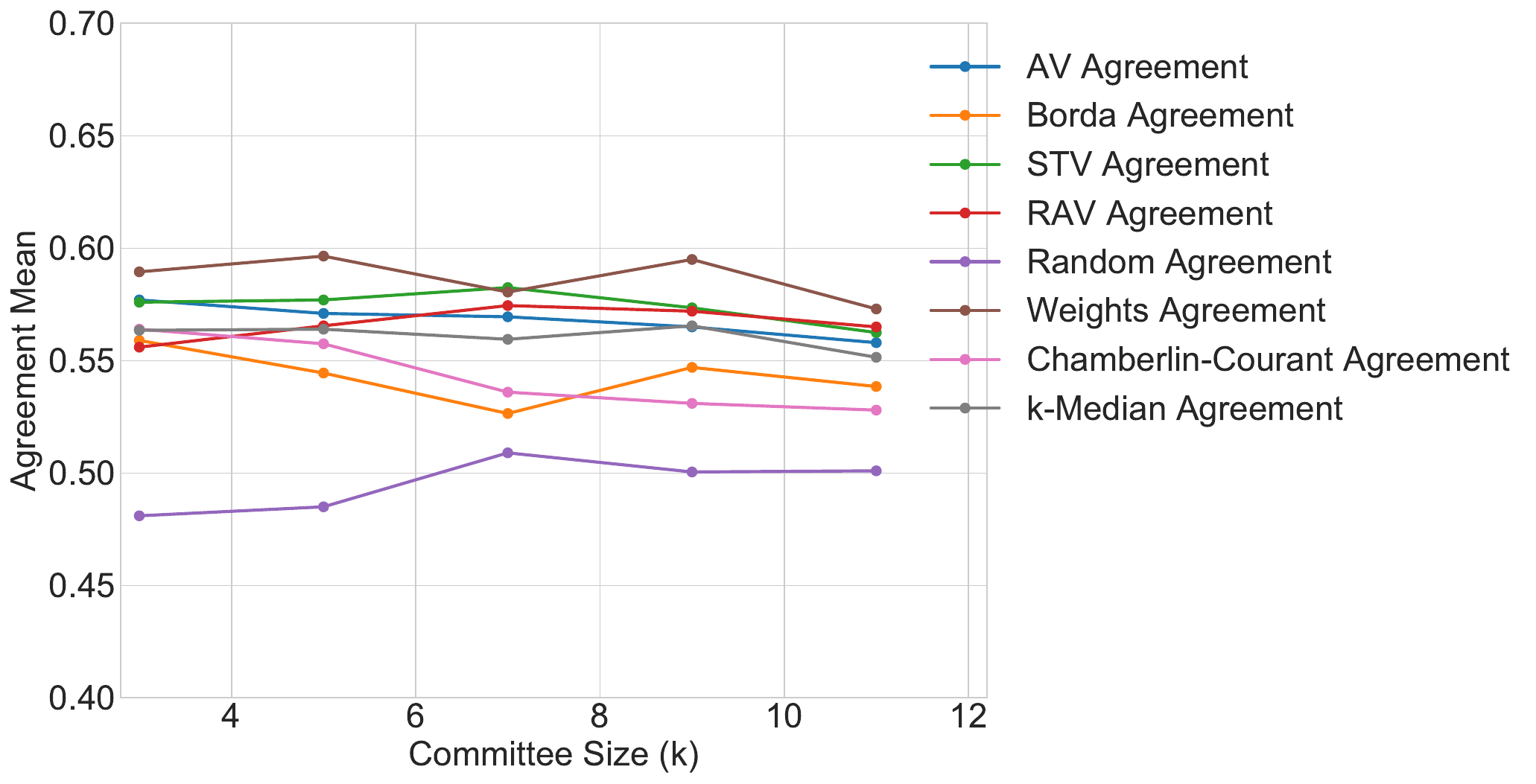}
    \caption{Comparison of NP-Hard rules with our polynomial rules for $|C|=17, |S|=80, |V|=51$.  We cannot scale this graph in the same was Figure \ref{fig:agree} due to the high computational cost of computing the winning sets for $k-$Median and Chamberlin-Courant.  However, from this small sample we see that Weights, STV, and AV all outperform Chamberlin-Courant and $k-$Median in terms of agreement.}
    \label{fig:np_vary_k}
\end{figure}

% \begin{itemize}
%     \item Show that for threshold approval prefs the best threshold is 0.5.
%     \item Simulation of DD with partial participation. Include variances to show it is very sensitive. The number of voters who support one issue minus number of voters who support the other is modeled by a random walk, so if there are 100 voters, the majority will contain no more than 55 voters with high probability ($55-45 = \sqrt{100}$)
    %
    % \item Later: Examine Direct Democracy with partial participation (compare variance to FRD?)
    % \item Later: Comparison with Liquid democracy
    % \begin{itemize}
    %     \item No delegation cycles
    %     \item No coercion-subversion dilemma (privacy vs. transparency)
    %     \item Fractional delegations
    %     \item Incentive to inform the public
    %     \item Conducive to writing legislation
    %     \item Distribution of power is public and easy to comprehend
    %     \item Default over representatives
    %     \item Can delegate on topics or for some amount of time
    %     \item Incentive for public deliberation and campaigning
    % \end{itemize}
    % \item Later: Add top t election approvals
% \end{itemize}

\end{document}